\documentclass[epj]{svjour}
\usepackage[T1]{fontenc}
\usepackage[export]{adjustbox}

\usepackage{graphics,graphicx}
\usepackage{hyperref} 
\usepackage{amsmath,amssymb,amsfonts}
\usepackage{algorithm,algorithmic}
\usepackage[shortlabels]{enumitem}
\usepackage{color}
\usepackage[backend=bibtex, style=nature ,natbib=true, sorting=none]{biblatex}
\AtEveryBibitem{\clearfield{title}}
\AtEveryBibitem{\clearfield{doi}}
\AtEveryBibitem{\clearfield{pages}}
\AtEveryBibitem{\clearfield{publisher}}
\AtEveryBibitem{\clearfield{issn}}
\newcommand{\panda}{PANDA}
\newcommand{\bab}{\textsc{Babai+20}}

\usepackage{multirow}
\usepackage{booktabs}
\usepackage{subcaption}
\usepackage{float}

\begin{document}

%
% \title{LOTF: a LOcal Track Finder algorithm for track reconstruction using the \panda\ Straw Tube Tracker}
\title{Reconstructing charged-particle trajectories in the \panda\ Straw Tube Tracker using the LOcal Track Finder (LOTF) algorithm. }

%\subtitle{}
\author{Simon Gazagnes\inst{1,2,3}\thanks{email: \emph{gazagnes@utexas.edu} (corresponding author)} \and Nasser Kalantar-Nayestanaki\inst{1} \and Johan G. Messchendorp\inst{1,4} \and Jenny Regina\inst{4} \and Tobias Stockmanns \inst{5}  \and Michael H.F. Wilkinson\inst{2}, on behalf of the PANDA collaboration
}                     % Do not remove
%
%\offprints{\emph{s.r.n.gazagnes@rug.nl}}          % Insert a name or remove this line
%
\titlerunning{LOTF: a LOcal Track Finder algorithm for the \panda\ STT}
\institute{Energy and Sustainability Research Institute Groningen, University of Groningen, 9700 CC Groningen, The Netherlands \and Bernoulli Institute for Mathematics, Computer Science and Artificial Intelligence, University of Groningen,  \\9700 AK Groningen, The Netherlands \and Department
of Astronomy, The University of Texas at Austin, 2515 Speedway, Stop C1400, Austin, TX 78712, USA \and GSI Helmholtzzentrum f\"ur Schwerionenforschung GmbH, 64291 Darmstadt, Germany \and 
Forschungszentrum J\"{u}lich, Institut f\"{u}r Kernphysik, 52428 J\"{u}lich, Germany}
\date{Received: date / Revised version: date}
\abstract{
We present the LOcal Track Finder (\textsc{lotf}) algorithm, a method that performs charged-particle trajectory reconstruction using the Straw Tube Tracker, one of the central trackers of the antiProton ANnihilation at DArmstadt (\panda) detector. The algorithm builds upon the neighboring relations of the tubes to connect individual hits and form track candidates. In addition, it uses a local fitting procedure to handle regions where several tracks overlap and utilizes a system of virtual nodes to reconstruct the z-information of the particle trajectories. We generated 30,000 events to assess the performance of our approach and compared its global track assignment efficiency with respect to two other track reconstruction methods. \textsc{lotf} has (1) an average of 85\% of \textit{found} tracks, (2) the largest number of \textit{Fully Pure} tracks, (3) the lowest amount of incorrect reconstructions, and (4) is significantly faster than the other two approaches. Further, we compared the z-reconstruction performance with one of the two alternative methods and show that \textsc{lotf} improves the median z-error by a factor of 8.7. Finally, we tested our method using 3,750 data sets composed of 4 events each, demonstrating that our approach handles cases in which events are mixed. The \textit{raw} (without parallelization) average reconstruction rate is about 68,000 hits/s, which makes the present algorithm promising for online data selection and processing.   } %end of abstract
\maketitle
\section{Introduction}
\label{intro}

Observing exotic particle states is important in characterizing the properties of the fundamental building blocks of matter. Producing unexpected or not yet observed particle states with a small cross-section requires a vast number of particle collisions to create a detectable signal above the background noise. This complex task is now achievable with modern particle detectors that operate at very high luminosities and interaction rates. The upcoming antiProton ANnihilation at DArmstadt experiment (\panda\footnote{\url{https://panda.gsi.de/}}, \cite{Ketzer:2006} \cite{barucca2021panda}), that will be installed at the Facility of Antiproton and Ion Research (FAIR\footnote{\url{https://fair-center.eu/}}, \cite{Durante:2019}), will provide such a setup. It will study interactions between protons or nuclei and antiprotons ($\overline{p}p$, $\overline{p}A$) using an antiproton beam with a momentum between 1.5 and 15 GeV/$c$, provided by the High Energy Storage Ring (HESR, \cite{hesr2020}) that impinges on a target of hydrogen or noble elements. The $\overline{p}p$ and $\overline{p}A$ collisions typically produce a wide variety of particles which makes them suited to explore \textit{e.g.} exotic particle states. \panda\ will work with two main operational modes: the high-resolution mode with interaction rates of up to 2 MHz in \cite{barucca2021panda}, and the high-luminosity mode with interaction rates of up to 20 MHz and data rates up to 200 Gigabytes per second \cite{pandaPhy2009}. Traditionally, particle-collision experiments use hardware triggers to reduce the data stream. However, this is not foreseen with \panda\ which will exploit software-based event filtering using online track reconstruction algorithms to select physically relevant events. Hence, efficient and fast track reconstruction algorithms are needed to identify the particle trajectories in a time frame matching the detector acquisition rate.

Most state-of-the-art track reconstruction algorithms are composed of two steps: a \textit{track finding} step, where the individual hits are grouped into track candidates using Pattern Recognition techniques, and a \textit{track fitting} step, where the particle properties are determined based on the tracks found. These two steps are often performed concurrently such that the particle properties (\textit{e.g.} their momentum) are estimated during the reconstruction. Several techniques exist to perform the \textit{track finding} step, usually categorized into \textit{local} or \textit{global} approaches. \textit{Local} approaches, such as the Kalman filter \cite{kalman1960}, reconstruct particle paths by iteratively connecting hits one by one. \textit{Global approaches}, such as the Hough transform \cite{Hough:1959qva}, tackle the reconstruction by determining all track candidates simultaneously using the entire set of detections at once.

% Recently, \citet{babai16,babai20} presented a hybrid approach building upon attribute-space-connected morphological filters \cite{wilkinson07:_attrib_space_connec_connec_filter} to reconstruct trajectories locally, based on the orientation of group of neighboring hits. Finally, the use of deep learning methods has also been explored to perform the track finding (\textit{e.g.}, \citet{esmail2019}).  \\

The performance of a given event reconstruction algorithm is usually measured in terms of (i) its \textit{efficiency}, \textit{i.e.} the fraction of real tracks that have been accurately identified; (ii) its \textit{purity}, \textit{i.e.} its ability to discard false hits or handle noisy data; and (iii) its computational speed. In practice, the design of the experiment itself must also be accounted for when implementing a track reconstruction algorithm. This is because specifics related to the efficiency of the data acquisition system and instrumental noise are also crucial aspects to consider. In general, all these aspects are tightly related to each other. For example, designing a very efficient track reconstruction procedure is often too slow to enable fast decision-making in real-time, and vice-versa. Hence, in-situ track reconstruction algorithms require an optimal balance between efficiency/purity and computational time, and must be optimized for the data acquisition it will work with.  

\begin{sloppypar}

In the context of the \panda\ experiment, several strategies were developed over the past years. These methods build upon techniques such as the Hough transformation \cite{bianchi2017,Alicke:2021omm}, the Recursive Annealing Fit \cite{Andersson2021AGA}, the cellular automaton with Riemann mapping \cite{Regina:2019dvn}, the triplet finder \cite{ADINETZ2014}, attribute-space-connected morphological filters \cite{babai16,babai20}, or deep-learning \cite{esmail2019, akram2022}. Additionally, a significant effort focused on parallelizing these algorithms using \textit{e.g.}, FPGAs \cite{Liang2017} or GPUs \cite{herten2015gpu, bianchi2017}. Finally, some algorithms focused on supplementing track finders, \textit{e.g.} the \textsc{PzFinder} \cite{Andersson:phd, Andersson2021AGA} that aims at extracting the longitudinal position of the charged particles.

In this paper, we present the LOcal Track Finder (\textsc{lotf}) algorithm, a novel method to perform track reconstruction using the Straw Tube Tracker (STT, \cite{pandastt}), an essential sub-detector for the track reconstruction and event identification of \panda. The focus of \textsc{lotf} is to enable an efficient in-situ track reconstruction. We developed a local approach using the neighboring relations of the tubes to connect individual hits and form track candidates (\textit{e.g.}, similar to \cite{ADINETZ2014, Regina:2019dvn, babai20}). Further, \textsc{lotf} builds upon a simple parametric fitting to disentangle individual trajectories in regions where several tracks overlap. Finally, it uses a system of virtual nodes, a technique originally proposed by \citet{babai20}, to reconstruct the z-information of the charged-particle trajectories. Its low computational complexity enables a fast and efficient track reconstruction, which makes it promising for online event selection. 

\end{sloppypar}

This paper is organized as follows: Section~\ref{sect:panda} presents the \panda\ central STT detector. Section~\ref{sect:alg} describes the implementation of our algorithm. Section~\ref{sect:perf} assesses the performance of our approach and Section~\ref{sect:conc} presents our conclusions and discusses future prospects. Appendix~\ref{app:app} details the main algorithms implemented in \textsc{lotf}.

\section{The \panda\ Straw Tube Tracker}
\label{sect:panda}

\begin{figure}
  \includegraphics[width=\hsize]{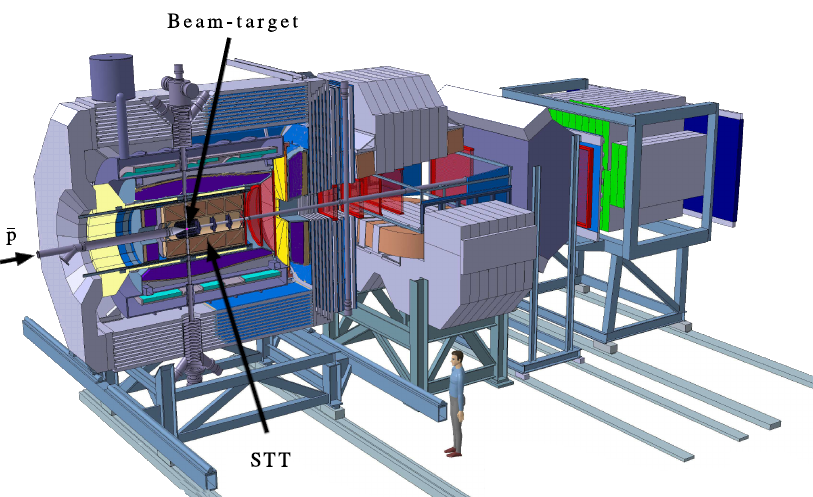}
    \includegraphics[width=\hsize]{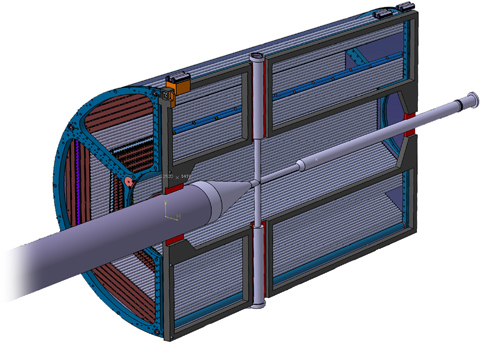}

    \caption{Top: a schematic representation of the  \panda\ detector system with the beam-target and STT highlighted, adapted from \cite{schwarz2012barrel}. In this work, we perform track reconstruction using the information from the STT, located around the beam-target interaction point. Bottom: a schematic representation of the STT, adapted from \cite{pandastt}.}
    \label{fig:panda}
\end{figure}

\begin{figure*}[!htbp]
\centering
    \resizebox{0.8\textwidth}{!}{%
    \includegraphics[trim={0 0 0 0},clip]{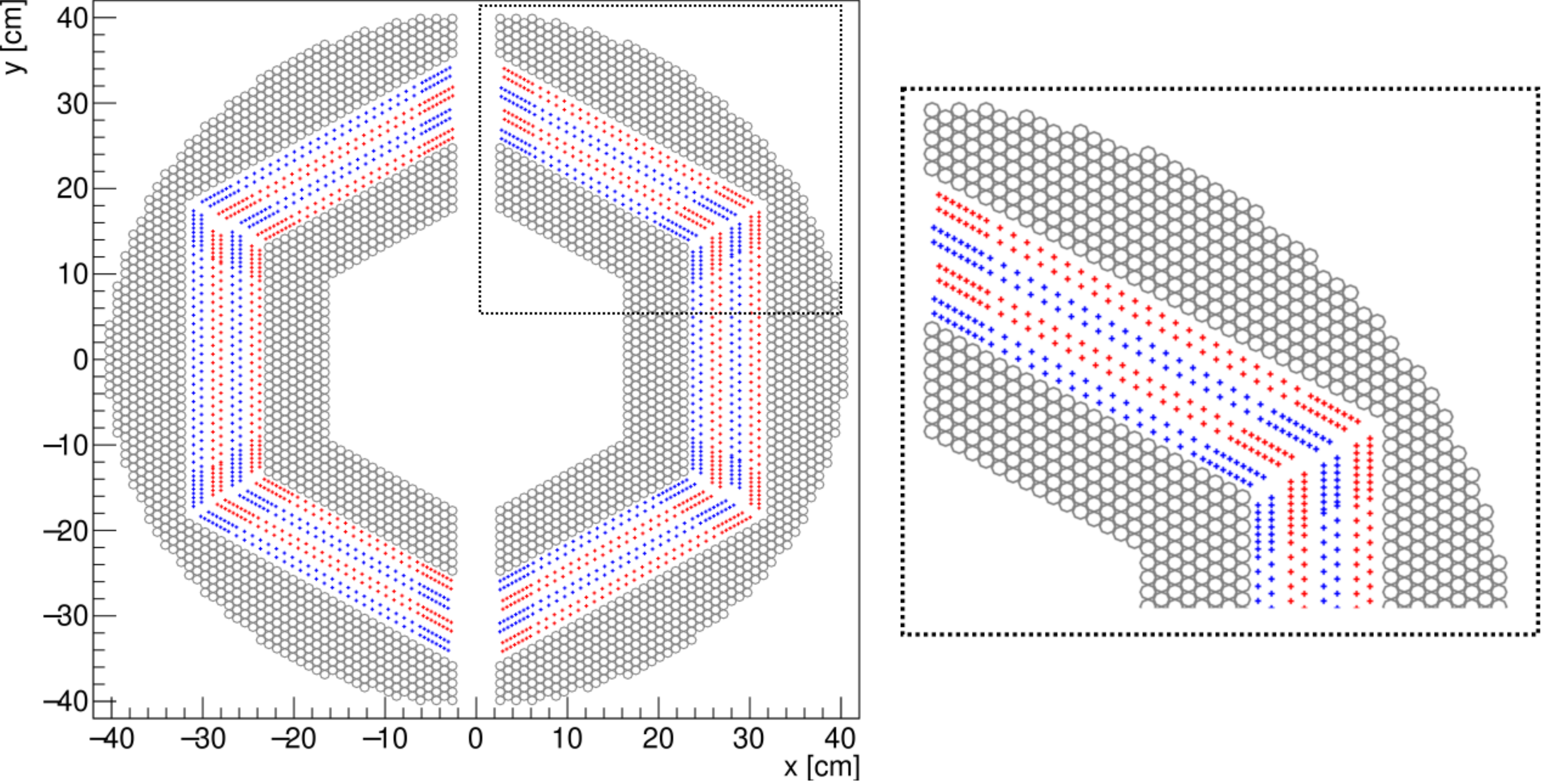}
    }
    
    \caption{The x-y projection of the tube positions in the STT (left), with a zoom-in on the top left section (right). The grey circles indicate the center-point coordinates of the axial tubes in the volume. The red and blue crosses show the center-point coordinates of the skewed tubes with an angle of $+2.9$ and $-2.9$ degrees, respectively.}
    \label{fig:xyproj}
\end{figure*}

\begin{figure*}[!htbp]
\centering
      \resizebox{0.43\textwidth}{!}{%
     \includegraphics{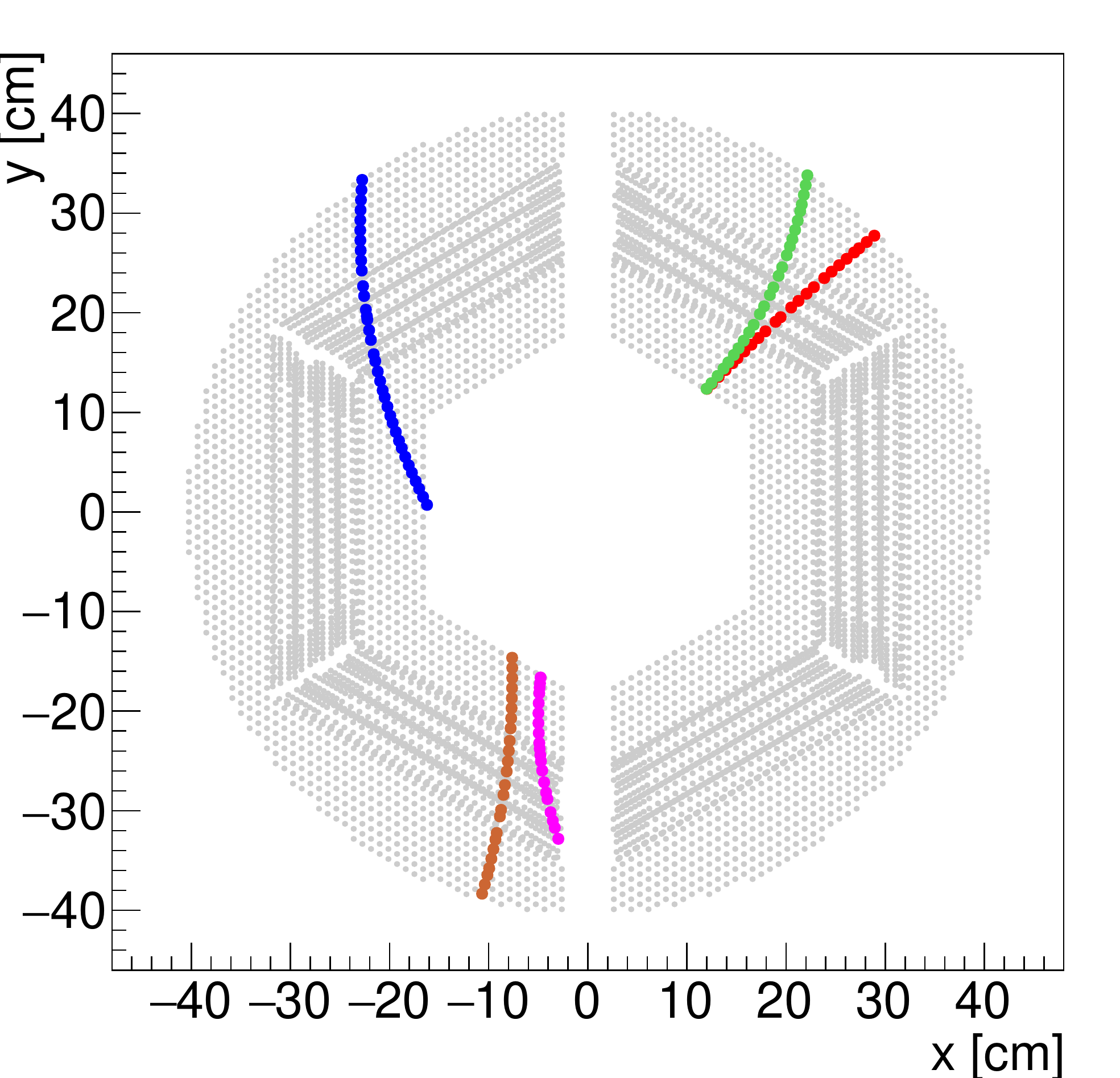}
    } 
     \resizebox{0.43\textwidth}{!}{%
    \includegraphics{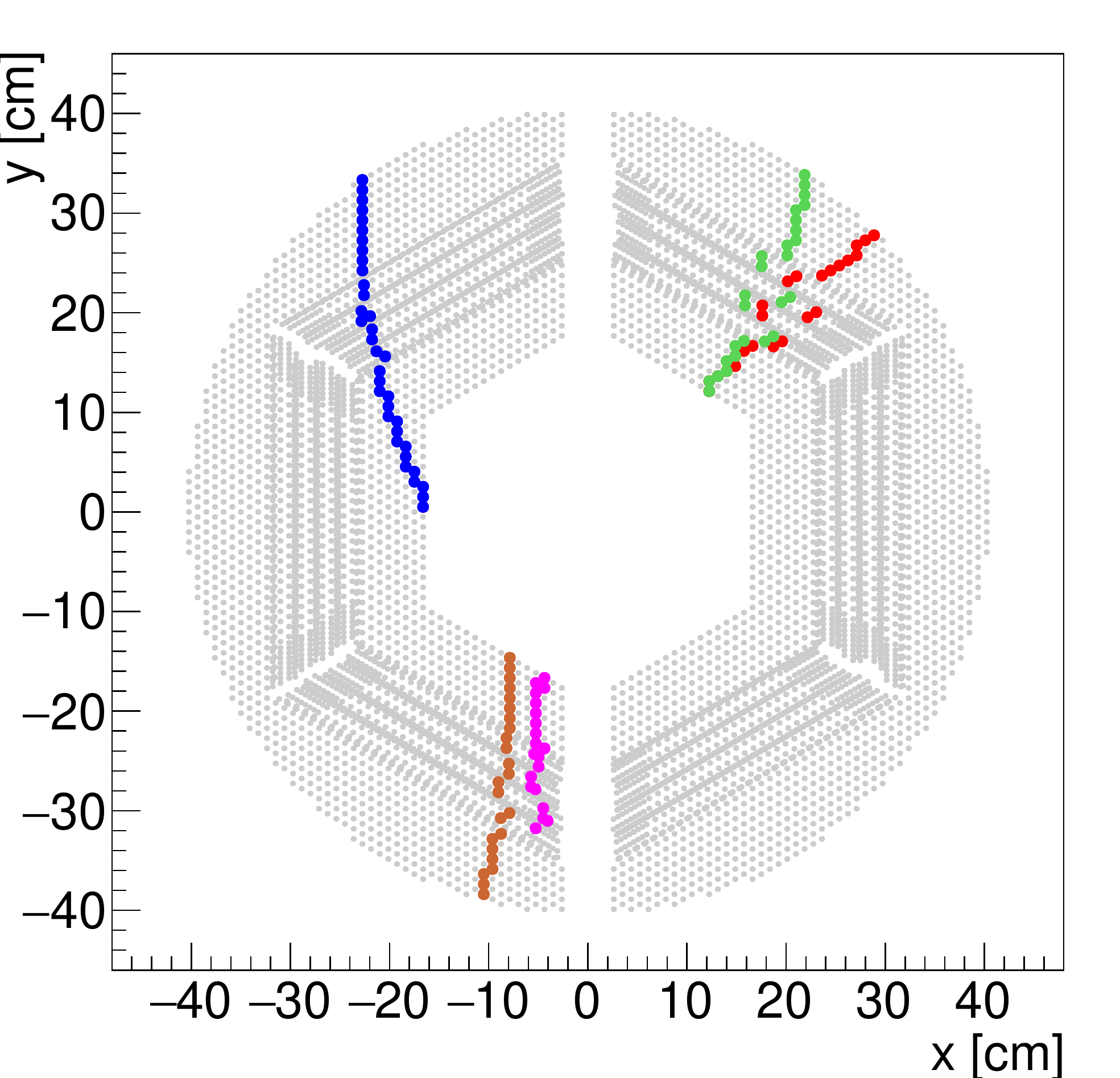}
    }
    \caption{Left: the x-y trajectories of 5 particles (in different colors) in a single event simulation. Right: the digitized version of the middle panel, showing the tubes that fired (\textit{i.e.} were hit) for each track. }
    \label{fig:xyproj_evt}
\end{figure*}

The \panda\ tracking system consists of central and forward trackers. The algorithm developed in this work performs track reconstruction using the STT system, which full description is provided in \cite{pandastt}. We detail here the most important characteristics.

\begin{sloppypar}
In \panda, the STT is a cylindrical volume located around the beam-target interaction point (see Figure~\ref{fig:panda}), such that the target beam, which has a full-width diameter of $\sim$3~mm, will pass through its center. The angular acceptance of the STT is from 20$^{\circ}$ to 140$^{\circ}$. Since \panda\ will use a central magnetic field of 2T generated by a superconducting solenoid, particles with a transverse momentum larger than 50 MeV will transit, partly or fully, through the STT.
\end{sloppypar}

Regarding its configuration, the STT cylindrical shape is characterized by an inner radius of 15 cm, an outer radius of 42 cm, and a length of 165 cm. Within this volume are arranged 4,224 straw tubes organized in 27 plane-parallel layers, with layer 0 being the closest to the beam-target interaction point. Each tube is 140 cm long with a 10 mm internal diameter and can be mounted in two different ways. In the innermost and outermost layers, the tubes are positioned parallel to the z-axis (beamline). In the eight middle layers, the straw tubes are skewed by $-2.9$ and $+2.9$ degrees with respect to the z-axis. The presence of these skewed tubes enables one to derive the z-information of the particle trajectory (see Section~\ref{sect:zpro}). 

Each straw tube is filled with an Ar/CO$_2$ gas mixture and contains a thin wire along its axis. An electric field is generated in the gas-filled area such that, when a charged particle passes through the tube, the gas becomes ionized. The gas ionization generates a current in the wire which is used as a readout signal (hereafter, we refer to these tubes as \textit{activated} or \textit{hit}). The time it takes the electrons to reach the wire is called the drift time. In general, the drift time is used to infer the \textit{isochrone radius},  \textit{i.e.}, the radius of the cylindrical surface that represents the possible positions that the particle might have traversed along the tube. Then, the exact particle path can be reconstructed by fitting a trajectory that is tangential to all the isochrone radii (see, \textit{e.g.}, Figure 3 in \cite{Andersson2021AGA}). The track reconstruction algorithm presented in this work does not use drift time information. This choice is motivated by two aspects. First, recovering the particle hit positions along each tube is not vital for identifying track candidates because one can simply group hits based on their neighborhood relations. The reconstruction of the exact particle trajectory can be performed later when one needs to extract tighter constraints on the particle momentum. Second, accounting for the isochrone radii involves an additional computational cost that can be prohibitive for online track reconstruction.

We must note that the drift time has important global implications for the track reconstruction in \panda. If the event rate is smaller than the maximal drift time of the straw tubes ($\approx 250$ ns), hits belonging to successive events can overlap during the readout of the STT data, which is referred to as event-mixing. Event-mixing has a relatively small influence at 2 MHz (1 event every 500 ns) but has strong repercussions for track reconstructions based on the high-luminosity mode (20 MHz, 1 event every 50 ns).  We discuss this aspect further in Section~\ref{sect:phaseII}.
%We discuss further this point in Section~\ref{sect:phaseII}.

 In this work, we represent the STT volume using a graph where each node represents a single tube.  Each node is parameterized by the half-length, the three-dimensional direction vector, and the coordinates of the center point of the tube it represents. Additionally, these nodes contain the list of the tubes' direct neighbors whose number can vary between 2 and 22 (skewed tubes or axial tubes on the edges of the STT volume have a varying number of neighbors). In the following sections, we present successively the x-y and z-y projections of the STT volume.

\subsection{x-y projection}
\label{sect:xypro}

Figure~\ref{fig:xyproj} shows the x-y projection of the STT volume based on the center-point coordinates of all the tubes. The tubes in the intermediate layers that have a skew angle with respect to the beam of $+2.9$ and $-2.9$ degrees are indicated with red and blue crosses, respectively. Figure~\ref{fig:xyproj} present a simulated event with five particle trajectories projected on the x-y plane of the STT. To generate this event, we used the \textsc{PandaRoot} software developed by the \panda\ collaboration \cite{Spataro2011} which contains a simulation and digitization package. We used the FRITIOF (FTF) generator with an antiproton beam momentum of 1.5 GeV/$c$. The left panel shows the simulated trajectories of the five charged particles resulting from a $\overline{p}p$ interaction, according to the transport model GEANT4 \cite{AGOSTINELLI2003,Allison2006, ALLISON2016}. The right panel shows the digitization  of the particles traveling along these paths based on the center-point coordinates of the tubes that are activated for each particle path. The exact hit positions along the skewed tubes can vary by $\pm$3 cm compared to the center-point coordinates. This causes the x-y projection of these digitized tracks to look discontinuous around the intermediate layers (\textit{i.e.}, the green and red tracks in the right panel).

An accurate track reconstruction in the x-y plane is crucial to determine the trajectory curvature and reconstructing the transverse momentum component of a particle. In the next section, we show the STT coordinates projected on the z-y plane, which is used to reconstruct the z-information of the particle tracks and derive the longitudinal momentum component.

\subsection{z projection}
\label{sect:zpro}

\begin{figure*}[!htbp]
    \centering
   % \vspace*{-0.5cm}
    % \begin{tabular}{ccc}
    %   \includegraphics[width = 0.28\textwidth,valign=t]{Images/virtualnodes.png}\hspace{0.1cm}  & \includegraphics[width = 0.34\textwidth,valign=t,trim={0 0 0cm 0.9cm},clip]{Images/STTGridVirt_XYPlane.pdf} &\includegraphics[width = 0.34\textwidth,valign=t,trim={0 0 0.cm 0.9cm},clip]{Images/STTGridVirt_ZYPlane.pdf} \\
    % \end{tabular}
    \resizebox{0.4\textwidth}{!}{%
\includegraphics[width = 0.3\textwidth,valign=t]{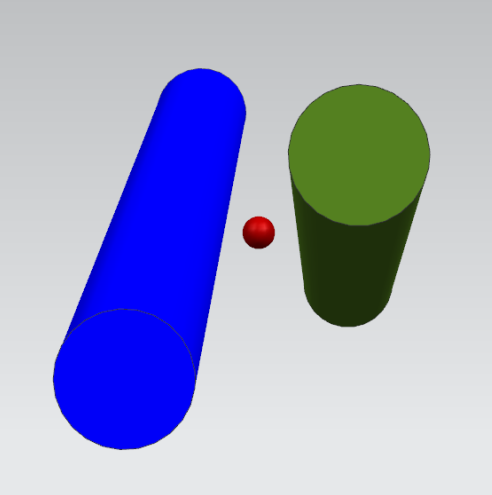}
} \\ \vspace{0.7cm}
    
      \resizebox{0.46\textwidth}{!}{%
\includegraphics[width = 0.34\textwidth,valign=t,trim={0 0 0cm 0.9cm},clip]{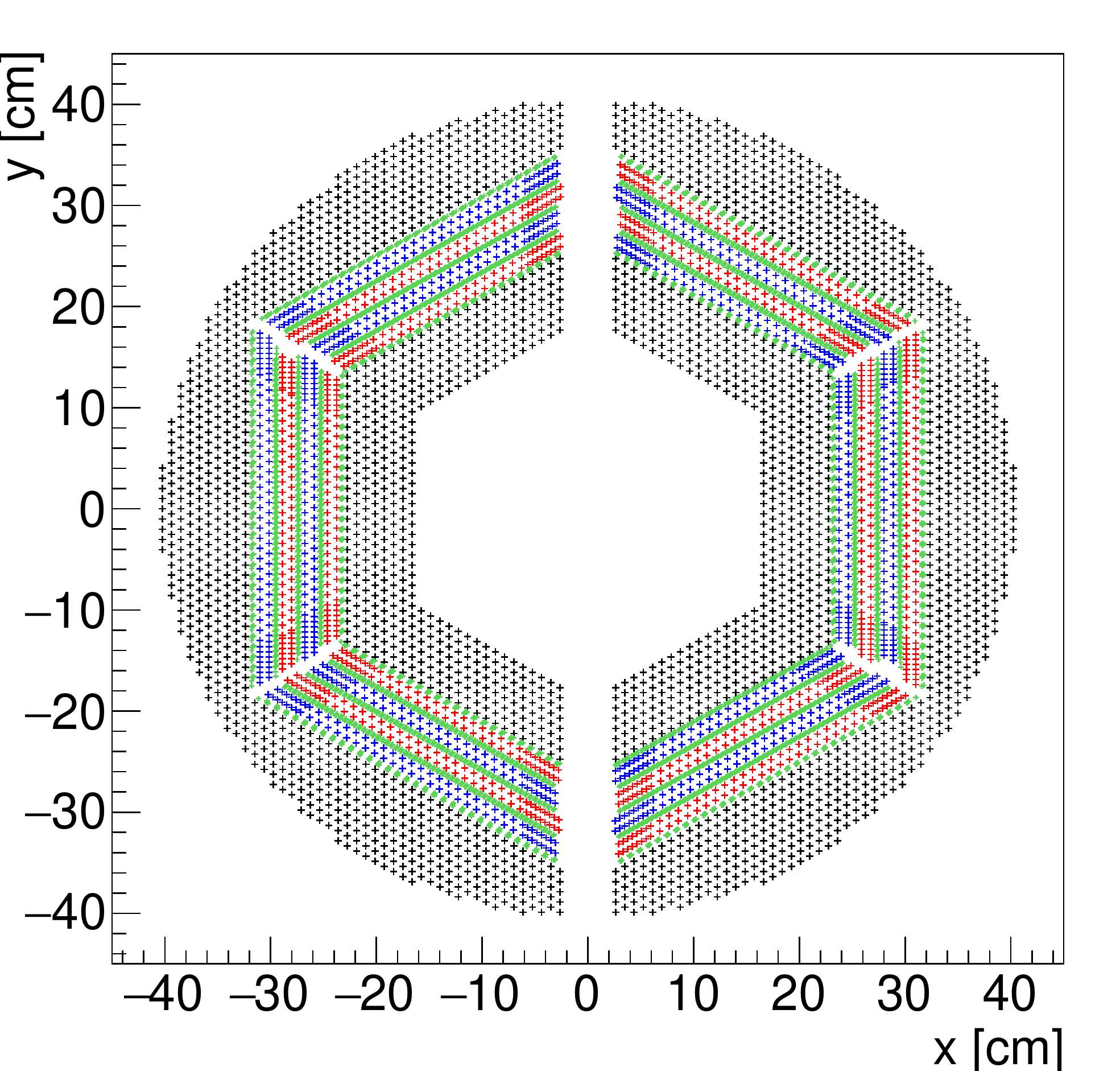}
} 
     \resizebox{0.46\textwidth}{!}{%
\includegraphics[width = 0.34\textwidth,valign=t,trim={0 0 0.cm 0.9cm},clip]{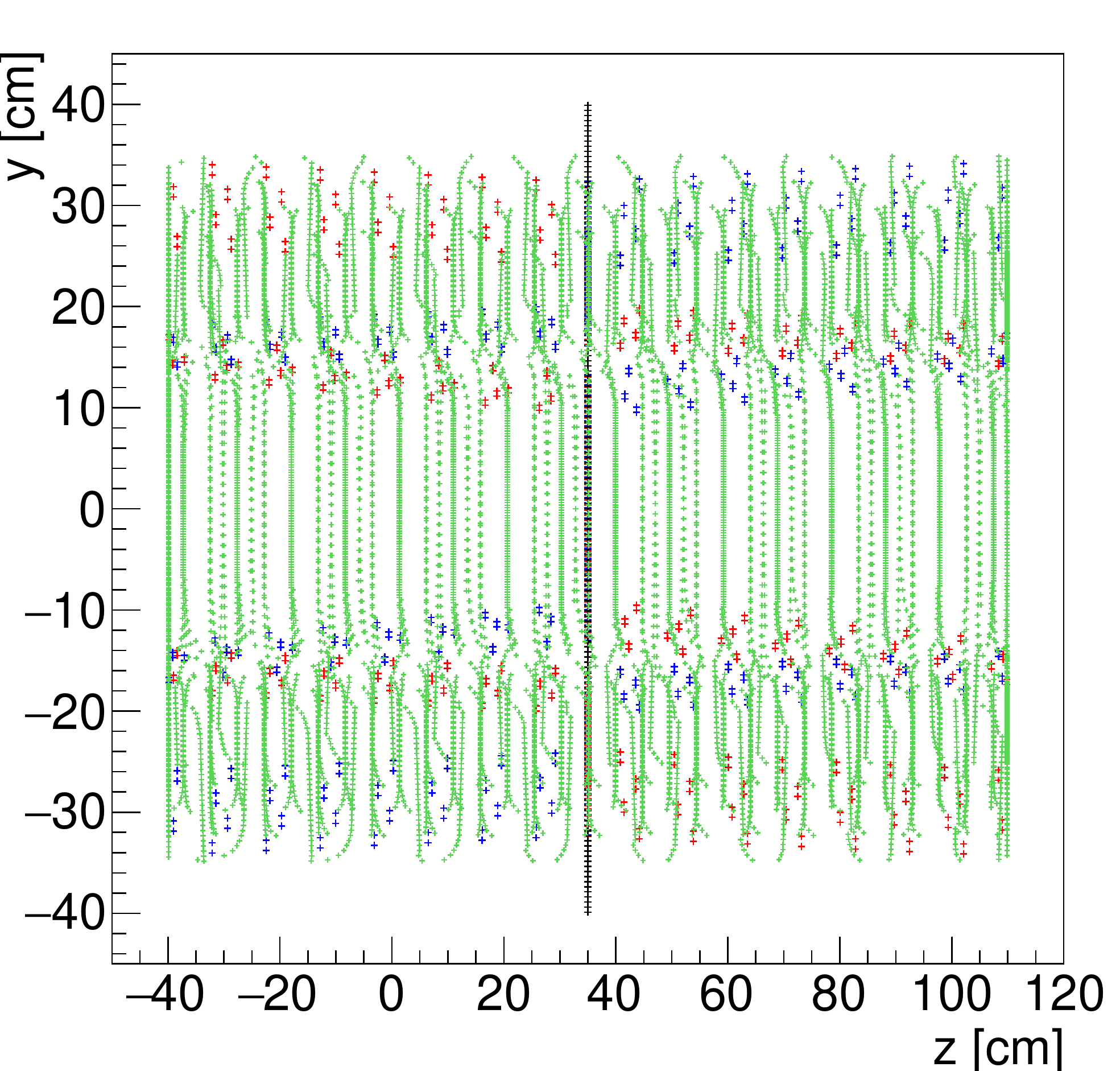}
}
    \caption{Top: a schematic view of the position of a virtual node (in red) for a pair of tubes having different z-slopes (Figure from \cite{babai16}). Bottom left: the x-y projection of the extended grid. Black crosses indicate the center-point coordinates of the axial tubes, and red and blue crosses show the center-point coordinates of the skewed tubes with an angle of $+2.9$ and $-2.9$ degrees, respectively. The virtual nodes appear in green. Bottom right: same but for the z-y projection. Because almost all the straw and skewed tubes span over the entire longitudinal length of the STT, their z center-point position is at z = 35~cm (the beam-interaction point is at z = 0). A small number of skewed tubes, located at the sector boundaries, have a shorter length, hence, they have a different z-coordinate.}
    \label{fig:extgrid}
\end{figure*}

\begin{figure*}[!htbp]
  \centering
    \includegraphics[width =  0.38\textwidth]{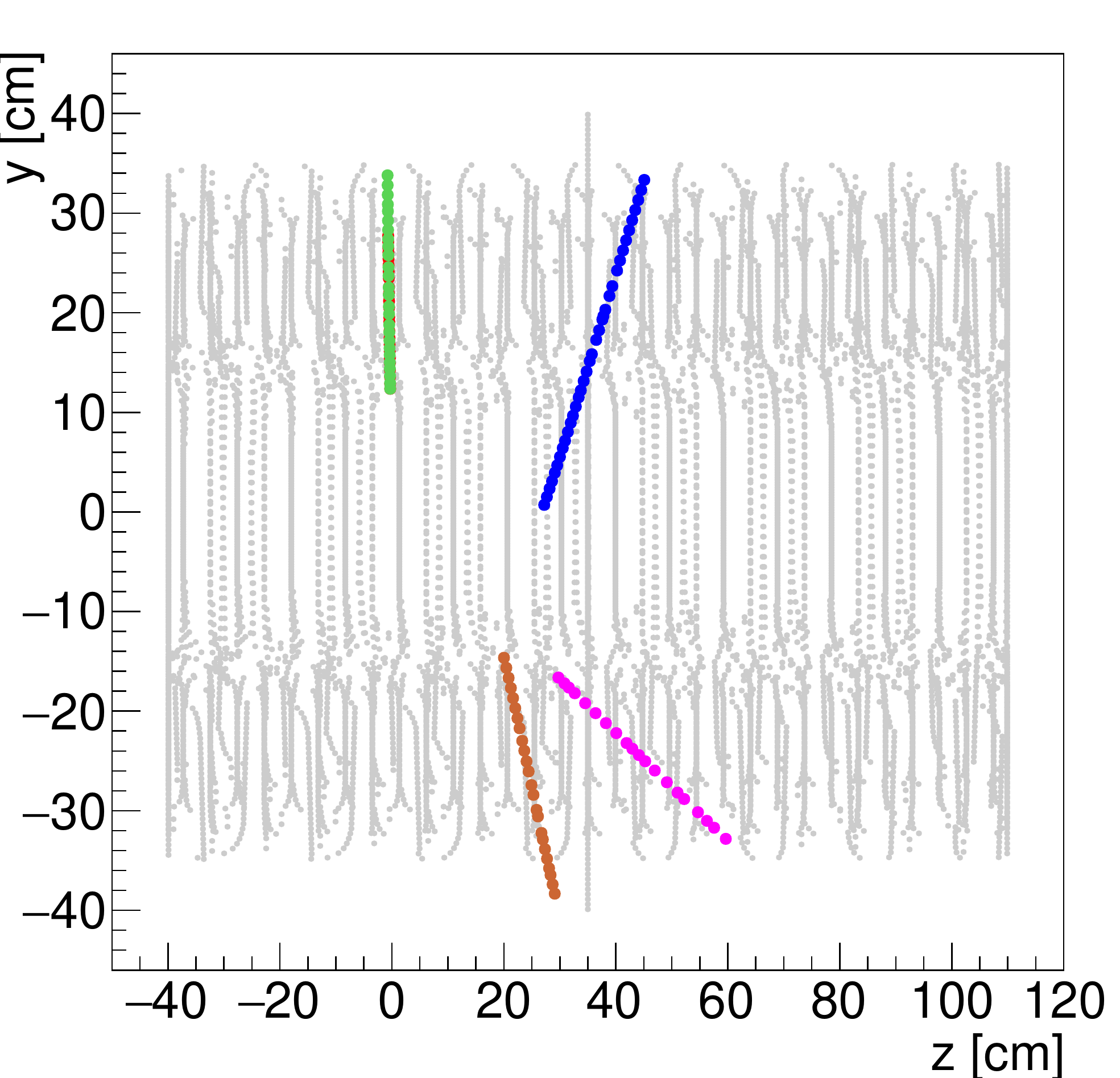} 
    \includegraphics[width =  0.38\textwidth]{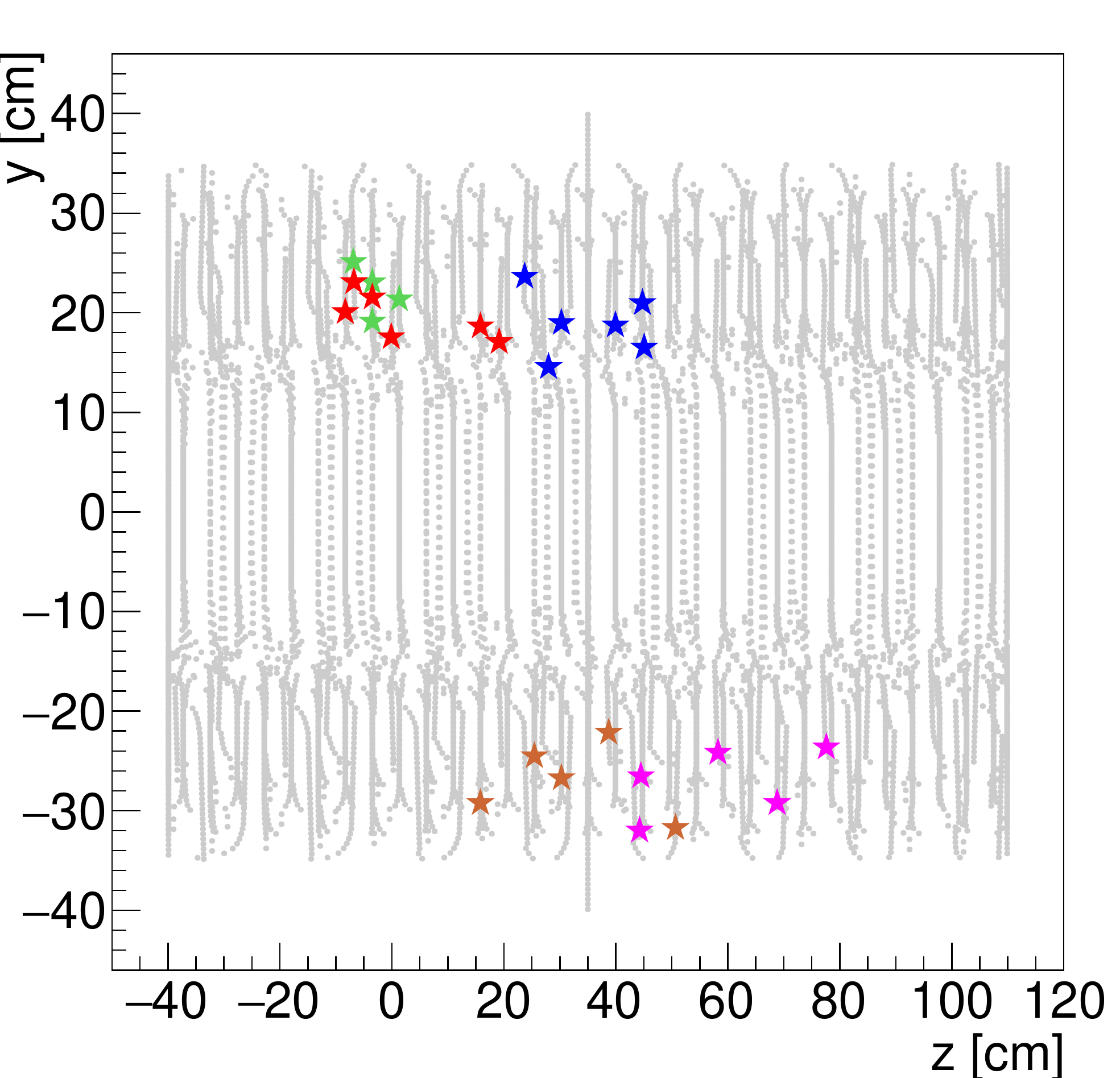} 
    \caption{Left: the simulated paths in the z-y plane for the five charged particles in the event shown in Section~\ref{sect:xypro}. The red and green tracks are overlapping in the upper left part. Right: the coordinates of the virtual nodes (stars) that are \textit{activated} for each track. }
    \label{fig:virtex}
\end{figure*}

 The reconstruction of the particle tracks in the z direction is much more complex than in the x-y plane. This is because, in the global coordinate system of the detector, the z center-point position of almost all the straw and skewed tubes is at z = 35~cm  (the interaction point is at x,y,z = (0,0,0)). A small number of skewed tubes, located at the sector boundaries, have a shorter length, hence they have a different z-coordinate compared to the full-length tubes. 
 
  For any activated tube, we do not have information about the exact z-coordinate of the particle hit. To refine the amount of information available in the z direction and improve the reconstruction of the z-information, we use a system of \textit{virtual nodes}, a technique introduced in \citet{babai16}. The concept of virtual nodes builds upon the design of the STT which includes layers of tubes having different slopes in the z direction. For two neighboring tubes having different z-slopes, there exists a \textit{virtual} intersection point in the three-dimensional space that corresponds to the point where the distance between the two tubes is minimal. This virtual point is represented schematically in the top panel of Figure~\ref{fig:extgrid}. The x, y, and z coordinates of this virtual point are strictly defined such that, if these two tubes are activated, one can assume that the particle passed through this virtual intersection point. This assumption is not always correct as it depends on the particle incident trajectory, yet it provides a first estimate to recover the particle z-information.
 
 The procedure to determine the x, y, and z coordinates of the \textit{virtual} nodes is the following: for all pairs of neighboring tubes, we extract the direction vectors of the two tubes and find the unit vector perpendicular to both lines. We then determine the point on this perpendicular line that is equidistant to both tubes. We take the coordinates of this point as the coordinates of the virtual node for this particular pair of tubes. All virtual nodes are set with only two neighbors which are the pair of tubes used to derive their coordinates. A virtual node is set as activated if both these tubes are activated. 

% Figure~\ref{fig:zyproj} shows the center-point coordinates of the straw tubes projected in the z-y plane. The black crosses indicate axial tubes, and red and blue crosses show the skewed tubes with an angle of $+2.9$ and $-2.9$ degrees, respectively.  Skewed tubes located at the sector boundaries have a smaller length, hence they have a different coordinate in z compared to the full-length tubes.

The bottom panels of Figure~\ref{fig:extgrid} present the extended grid, in the x-y (left panel) and z-y (right panel) planes, including the virtual nodes in green. For the x-y plane, virtual nodes span in the inter-layers between the axial, $+2.9$, and $-2.9$ degrees skewed tubes. For the z-y plane, these nodes substantially help to refine the discretization \footnote{In all this work, we only use the x-y and z-y plane to visualize the track reconstruction.}. Hence, they provide key information for the reconstruction of the particle trajectories in the z-direction. We must note that, while these virtual nodes were already introduced by \citet{babai16, babai20}, the authors used a different approach to compute their coordinates. For each pair of tubes with different slopes, the authors defined a two-dimensional plane using the central-point x-y coordinates of each tube. Then, they computed the virtual point coordinates as the coordinates of the point at the center of this two-dimensional plane. The positions of the virtual nodes using their approach significantly differ from our method which assumes that the particle passed through the point where the distance between both tubes is minimal. As a consequence, the z-coordinates of the virtual nodes in \cite{babai20} lie in a narrower range, and this has important shortcomings for the z-reconstruction. In Section~\ref{sect:xyrec}, we show that our approach provides a better reconstruction of the z-information.

The z-y projection of the particle tracks shown in Figure~\ref{fig:xyproj_evt} is presented in Figure~\ref{fig:virtex}. The left panel presents the simulated trajectories of the five particles in the z-y plane and the right panel displays the z-y coordinates of the virtual nodes (stars) that are activated. The comparison of the two panels illustrates that virtual nodes only provide a first estimate for the particle z-transit. Section~\ref{sect:zrec} further details how we build upon these nodes to reconstruct tracks that resemble the simulated ones.  

\section{The \textsc{LOTF} algorithm}
\label{sect:alg}

\begin{sloppypar}

In this work, we aim at providing an algorithm suitable to perform an in-situ track reconstruction using the STT information, both in the high-resolution (2 MHz interaction rate) and high-luminosity (20 MHz interaction rate) operational modes. Performing a real-time reconstruction is challenging given the diversity of physical processes, involving complex trajectories' morphology with long-lived particles and tracks from secondary decay particles. Hence, the tracking system must have both high track reconstruction efficiency and a high reconstruction rate matching the data acquisition rate. To achieve these requirements, \textsc{LOTF} builds upon low-computational complexity techniques, selected for their potential in extracting complex, irregular, and potentially overlapping patterns in the STT. Additionally, while some track reconstruction methods perform the track reconstruction in a different parameter
space to (\textit{e.g.}, the Hough space \cite{bianchi2017} or orientation space \cite{babai20}), \textsc{lotf} only uses the Cartesian space (\textit{i.e.,} the x, y, and z coordinates of the different hits), hence avoiding potential overheads due to space transformations.

Section~\ref{sect:xyrec} details the x-y reconstruction strategy and Section~\ref{sect:zrec} describes the procedure to recover the z-information of the charged-particle trajectories.
\end{sloppypar}

\subsection{x-y reconstruction}
\label{sect:xyrec}

\begin{figure*}[!ht]
    \begin{tabular}{c | c | c | c | c}
 \hspace*{-6mm}   \includegraphics[height = 2.78cm]{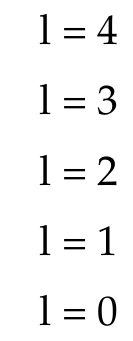}
 \hspace*{-3.5mm}   \includegraphics[height = 2.78cm]{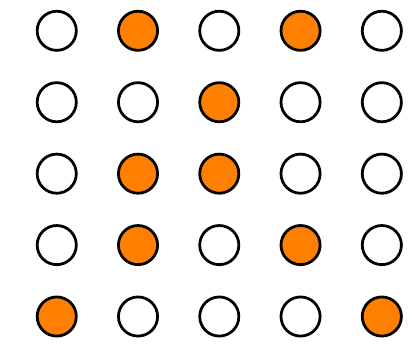}&
 \hspace*{-3mm}   \includegraphics[height = 2.78cm]{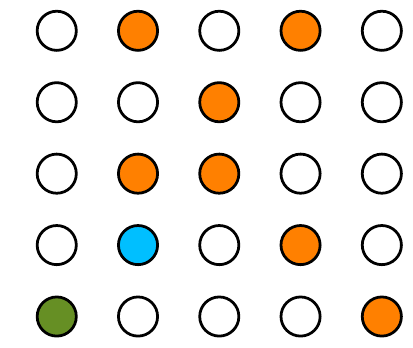} &
 \hspace*{-3mm}   \includegraphics[height = 2.78cm]{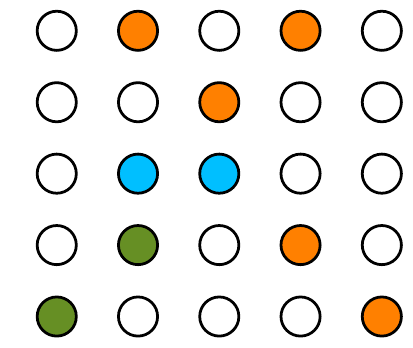} &
  \hspace*{-3mm}  \includegraphics[height = 2.78cm]{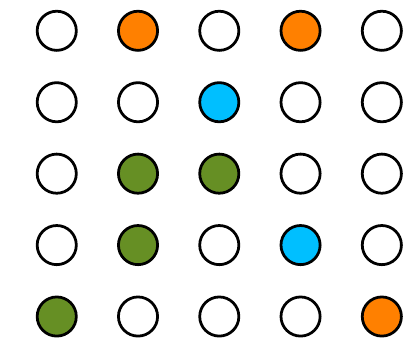} &
  \hspace*{-3mm}  \includegraphics[height = 2.78cm]{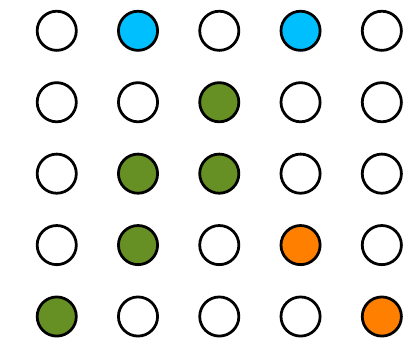} \\
(a)&  (b) & (c) & (d) & (e)
\end{tabular}
    \caption{A schematic example showing the functioning of the \textit{connect} procedure. In practice, our algorithm is applied to the STT hexagonal geometry where each tube can have up to 22 neighbors. In this schematic, for simplicity, we define a squared geometric configuration where each tube (empty black circle) can have at most 8 neighbors.} Orange circles are activated tubes, green circles are tubes that belong to the track being reconstructed, and blue circles show the current list of activated neighbors at each step. The left panel shows the index of each layer l.  In (a), we show the set of activated tubes. In (b), we start from a node in the inner layer (l = 0). This node has only one active neighbor, thus the case is trivial and we add this neighbor to the current track. In (c), we have two neighbors on layer 2. Both these nodes are adjacent and located consistently with respect to the current direction of the track, hence we connect both to the current track. In (d), we have neighbors both in the upper and lower layer. Given that we were adding nodes on increasing layer number (from l = 0 to l = 2), we connect only the node that belongs to layer 3. In (e), we encounter a complex case where we have two neighbors on the next layer, but these nodes are not adjacent. Finding which of these nodes should be added is not trivial. The \textit{connect} procedure stops and the reconstruction will continue during the \textit{fitting} phase. 
    \label{fig:conex}
\end{figure*}

Our approach to reconstructing the particle trajectories in the x-y plane consists of three steps:
\begin{itemize}
    \item The \textit{connect} phase identifies track candidates based on hits located in the inner- and outer-most  tube layers.
    \item The \textit{fitting} phase extends the reconstruction in regions where several track candidates overlap using a parametric fitting procedure.
    \item The \textit{merging} phase combines track candidates to produce the final reconstructed tracks.
\end{itemize}

\noindent We describe these phases in the following sub-sections.

\subsubsection{The \textit{connect} phase}
\label{sect:connect}

\begin{figure}[!htbp]
\centering
 \begin{tabular}{c | c}
    \hspace{-7mm} \includegraphics[height = 3.3cm]{Images/layers.pdf}
     \hspace{-4mm} \includegraphics[height = 3.3cm]{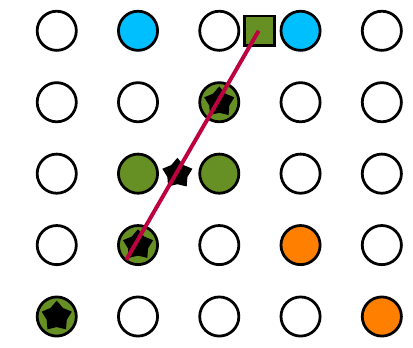} &
  \hspace{-4mm} \includegraphics[height = 3.3cm]{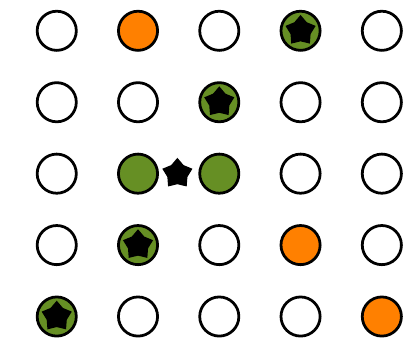} \\
     (a) & (b)
\end{tabular}
    \caption{The continuation of the schematic example presented in Figure~\ref{fig:conex}. Orange circles show activated tubes, green circles represent tubes that belong to the track being reconstructed, and blue circles show the current list of neighbors. The black stars show the position of the \textit{anchor} nodes for the green track. On layer l = 2, the anchor node position corresponds to the average of the coordinates of the two green nodes.  In (a), we fit a parametric line using the position of the three most recent anchors in the track. The green square shows the predicted position of the next hit given the line equation and the layer distance. The right blue node has the smallest euclidean distance to this prediction, hence it is connected to the track in (b). }
    \label{fig:fitex}
\end{figure}

Algorithm~\ref{alg:connect} in Appendix~\ref{app:connect} details the procedure used during the \textit{connect} phase, and we summarize the main steps here. Algorithm~\ref{alg:connect} aims at identifying portions of tracks that are isolated. To achieve this, we extract the activated nodes that have less than five neighbors (we consider as neighbors the activated tubes directly surrounding the current node) and that belong to the inner- and outer-most layers of the STT (we refer to these layers as the ``limit layers"). We note that the choice of using five neighbors as a threshold is empirical, but we find that it generally characterizes configurations where tracks are overlapping. Additionally, we extract nodes located in limit layers because they usually correspond to a track extremity. We start from these nodes and iteratively look among the surrounding neighbors to identify the track continuation. These neighbors are added to the current track if there is a low complexity in assessing whether they belong to the same track or not. We consider a case to be complex if the current node 
\begin{enumerate}
    \item has too many neighbors (more than 5),
    \item has neighbors in multiple directions (\textit{i.e.}, located on all surrounding layers),
    \item has no other activated neighbors surrounding it.
\end{enumerate}

We note that \#1 and \#2 usually suggest that there might be several adjacent or overlapping tracks around the region where the current node is located. These cases are complex and are resolved during the \textit{fitting} phase. On the other hand, one might encounter case \#3 if a track is complete or if there exists a gap such that the next neighbors are not direct neighbors of the last-added node. To disentangle both cases, we label the track as complete if it starts and ends in the limit layers of the STT. If this is not the case, we flag the track to be investigated during the \textit{merging phase}, as there might exist another track candidate it can be connected to.

The \textit{connect} procedure is sufficient to find  all ``simple" isolated tracks with no or few complex neighboring cases entirely. However, it does not perform a complete reconstruction for adjacent or overlapping tracks which are only partially found at the end of this phase. Figure~\ref{fig:conex} shows a schematic representation of a track reconstruction using the \textit{connect} procedure.

%\textbf{We note that, for simplicity, this example builds upon a simplified ``squared" geometric configuration where each node can have at most 8 neighbors. In practice, our algorithm is applied to the STT hexagonal geometry where each node can have up to 22 neighbors.}

\subsubsection{The \textit{fitting} phase}
\label{sect:fitting}

During this phase, we resume the reconstruction of partially reconstructed tracks. Algorithm~\ref{alg:fitting} in Appendix~\ref{app:fit}  details the fitting procedure and we present the main steps here. As mentioned in the previous section, the \textit{connect} procedure provides incomplete track portions in regions where multiple particle trajectories might cross or be too close to each other. In such configurations, finding the track continuation based on the list of neighbors around the last node added to the track is not trivial. We use a linear fitting approach based on a system of \textit{anchors} nodes to resolve these complex cases. \textit{Anchor} nodes are supplemental nodes representing at maximum three individual tubes on the same layer. Since several adjacent tubes on the same layer might be activated by the transit of a single particle, anchor nodes provide a simple way to join these multiple detections in a single \textit{anchor} (we define the size of an anchor node as the number of tubes it represents). 

Anchor nodes are determined on the fly during the track reconstruction. We construct a new anchor node each time a node from a different layer than the previous node is added to the track. The coordinates of a given anchor node are obtained by taking the average of the coordinates of the nodes it contains. In general, the \textit{anchor} system virtually smooths the current track candidate by only taking a smaller subset of nodes. Since the spatial distribution of these anchor nodes is smaller than for the individual tubes, they are well suited for the fitting procedure used to identify the best track continuation.

The fitting strategy is implemented as follows: we use the x-y coordinates of the three closest anchor nodes (at most) to the track extremity and fit a parametric line in the x-y plane, which equation is given by $x = x_0 +x_1*t$ and $y = y_0 + y_1*t$ with $t$ the independent variable. The choice of a linear fit might appear counter-intuitive since particles have circular motions, whose curvature depends on their momenta. Yet, as emphasized by, \textit{e.g.}, Figure~\ref{fig:xyproj_evt}, straight lines provide a fair approximation for the particle trajectories on small scales (three different layers at most).  Importantly, this approach has a small computational complexity which makes it promising to minimize the overall computational time. We note that we also tested a quadratic fitting approach, and found that the computational overhead was negligible when using a limited number of points. Nevertheless, we did not find significant improvements in the fit quality, and this approach did not perform well for short track candidates (less than three anchor nodes, hence not enough information to constrain the fit), or when applied to low-momentum particles circling within the STT volume. Therefore, we use the linear fitting approach as the default option in \textsc{lotf}, but the current code implementation enables the user to switch to a quadratic fit if preferred.

The resulting line equations enable us to predict the position of the next hit based on the current portion of the track identified. Using this prediction, we look among all potential neighbors to find the one closest to it. We add a fixed threshold of 5 cm to the maximal distance a node can have compared to the predicted position. 5 cm is roughly two times the maximum distance that can be found between the center point of two neighboring tubes. Hence, using this threshold avoids adding nodes unrealistically far from the current track extremity.   If a neighboring node successfully meets this criterion, and if it was not already assigned to another track candidate, it is added to the reconstructed track. On the other hand, if the node was already connected to another track candidate, we test whether both tracks should be merged according to their spatial orientation. To achieve this, we derive a local direction vector around the track extremities such that both tracks are flagged to be merged if the angle between their direction vectors is larger than 110 degrees (180 degrees means that both vectors are parallel). We note that this criterion has been chosen empirically as the value that works best after a visual inspection of a few events. Using a higher (lower) value decreases (increases) the likeliness that two tracks are merged. In general, we find that varying this criterion by $\pm 20$ degrees has a limited impact on the reconstruction performance of event-based data with low track multiplicity because the number of complex merging cases (\textit{i.e.}, cases where a single track candidate could be merged to different tracklets) remains small. Nevertheless, we have not extensively tested the limit of this value in the context of event-mixing data sets where the number of overlapping particle trajectories can be much larger (see Section~\ref{sect:phaseII}). We plan to work on optimizing this criterion in future developments focused on improving the high-luminosity track reconstruction performance.

For any track, the fitting procedure ends if (i) there are no further neighbors; (ii) no good neighbor is matching the predicted position of the next hit; or (iii) we found a neighboring track to connect with the current one.

\begin{sloppypar}
Figure~\ref{fig:fitex} illustrates the fitting procedure using the schematic configuration shown in Figure~\ref{fig:conex}.
\end{sloppypar}

\subsubsection{The \textit{merging} phase}
\label{sect:mergingg}

Before merging the tracks that were flagged during the \textit{fitting} phase, we perform a global check of all tracks to identify particle trajectories that might still be incomplete. As specified in Section~\ref{sect:connect}, we assume that a track is complete if its two extremities are on the boundaries of the STT. Incomplete tracks, as defined using this criterion, are either tracks whose extremities lie in the intermediate layers of the STT or tracks having a hit gap such that the remaining portion of the track to be connected is not directly neighboring the current track extremity. In the latter scenario, we typically reconstruct several track candidates that correspond to a single simulated trajectory. To determine whether a track candidate should be extended despite not having direct neighbors, we look for other track candidates within a radius of 5 cm around its extremities. As noted in Section~\ref{sect:fitting}, this distance represents a gap of at least two tubes between the two track extremities. If we find one neighboring track within this radius, and if its direction vector is consistent with the current track (using the same criterion as detailed previously, \textit{i.e.}, the angle between their direction vectors is larger than 110 degrees), both are flagged to be merged.

Once we tested all tracks, for each pair of tracks to be merged, their respective nodes are combined consistently into a new path candidate. 

\subsubsection{Example of x-y reconstructions}

 \begin{figure*}[!htbp]
 \centering
    \includegraphics[width = 0.32\hsize]{Images/evt0_simuxy.pdf} 
    \includegraphics[width = 0.32\hsize]{Images/evt0_read.pdf} \\
    \includegraphics[width = 0.32\hsize]{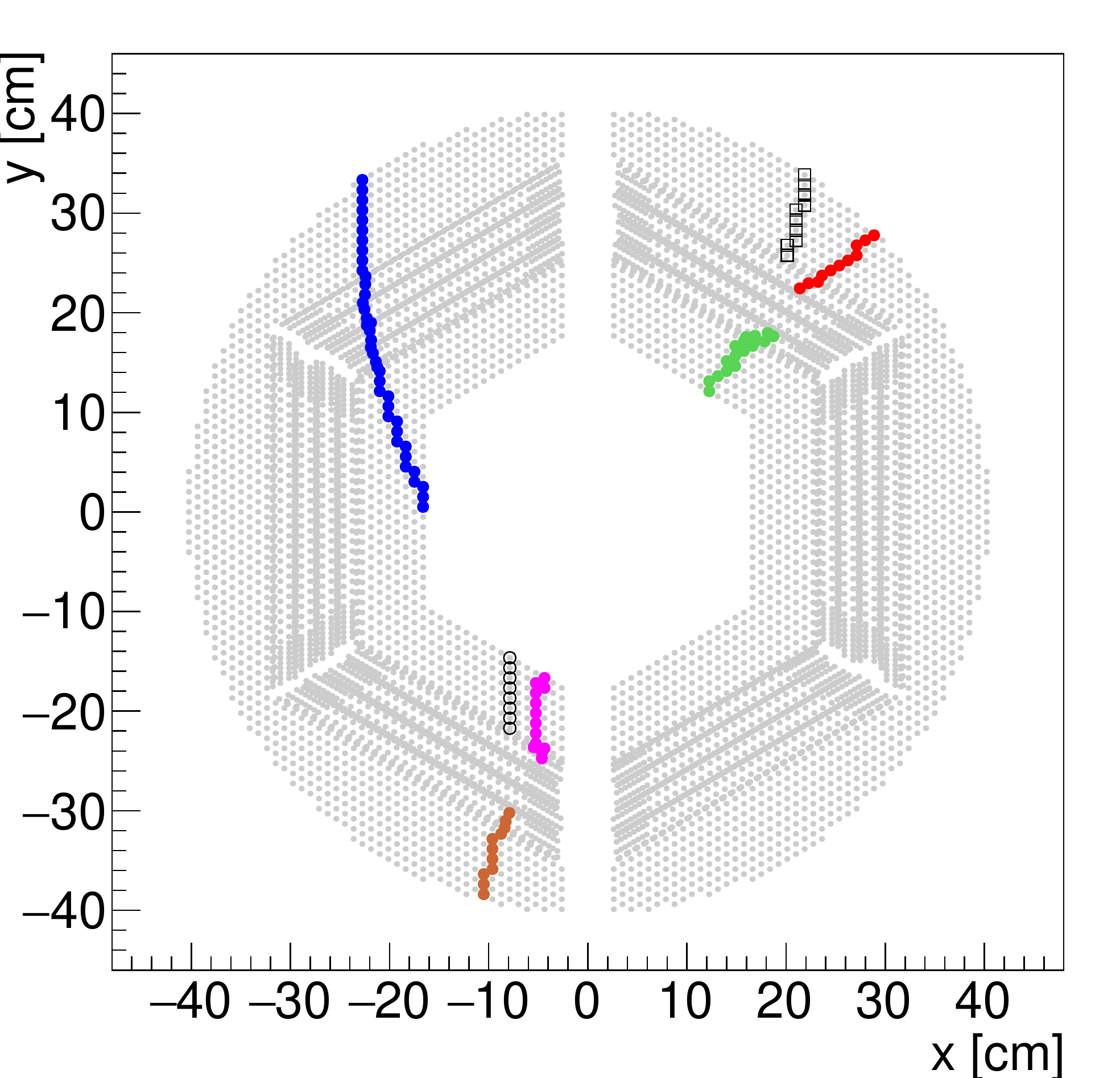} 
    \includegraphics[width =0.32\hsize]{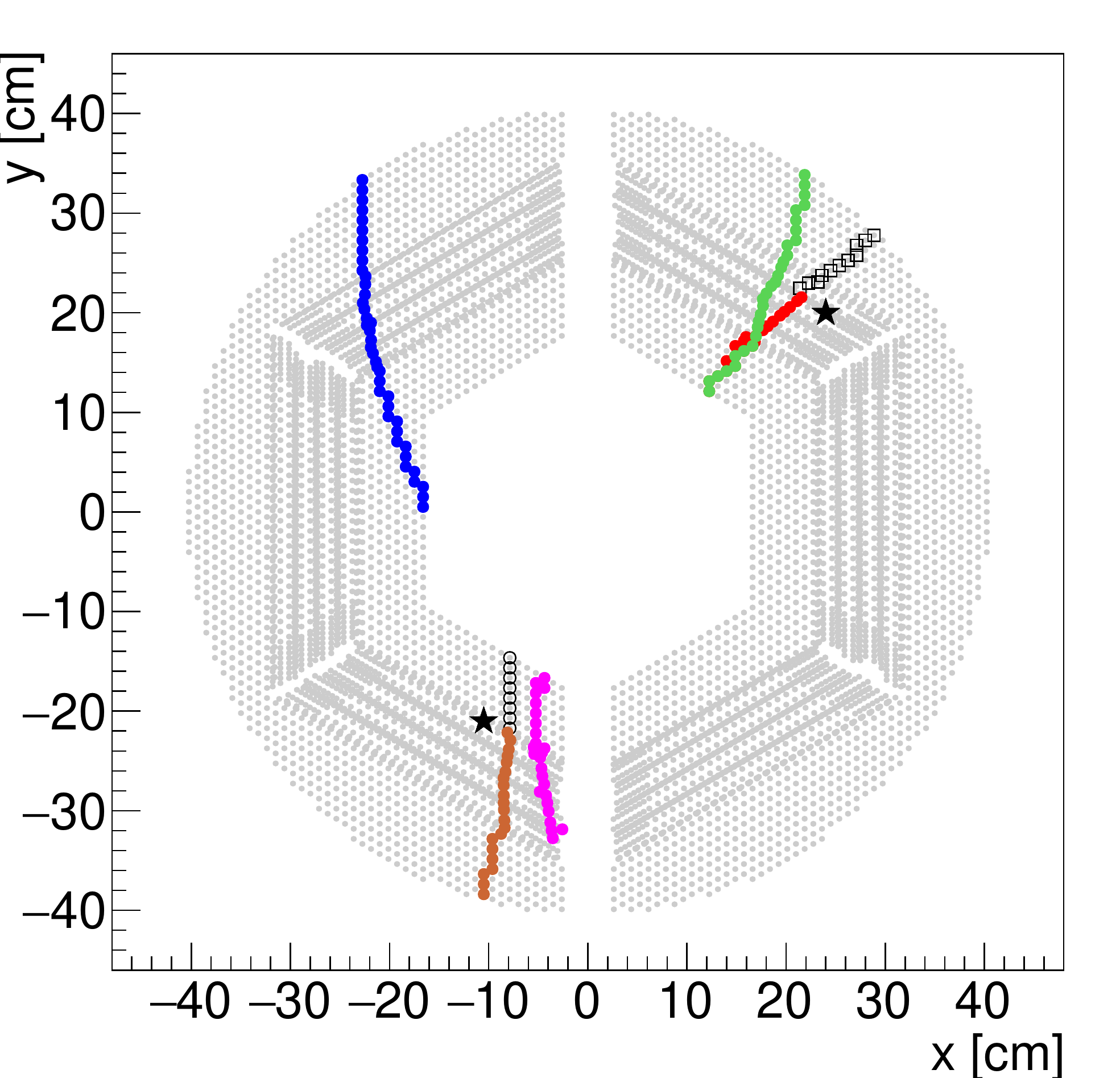} 
    \includegraphics[width = 0.32\hsize]{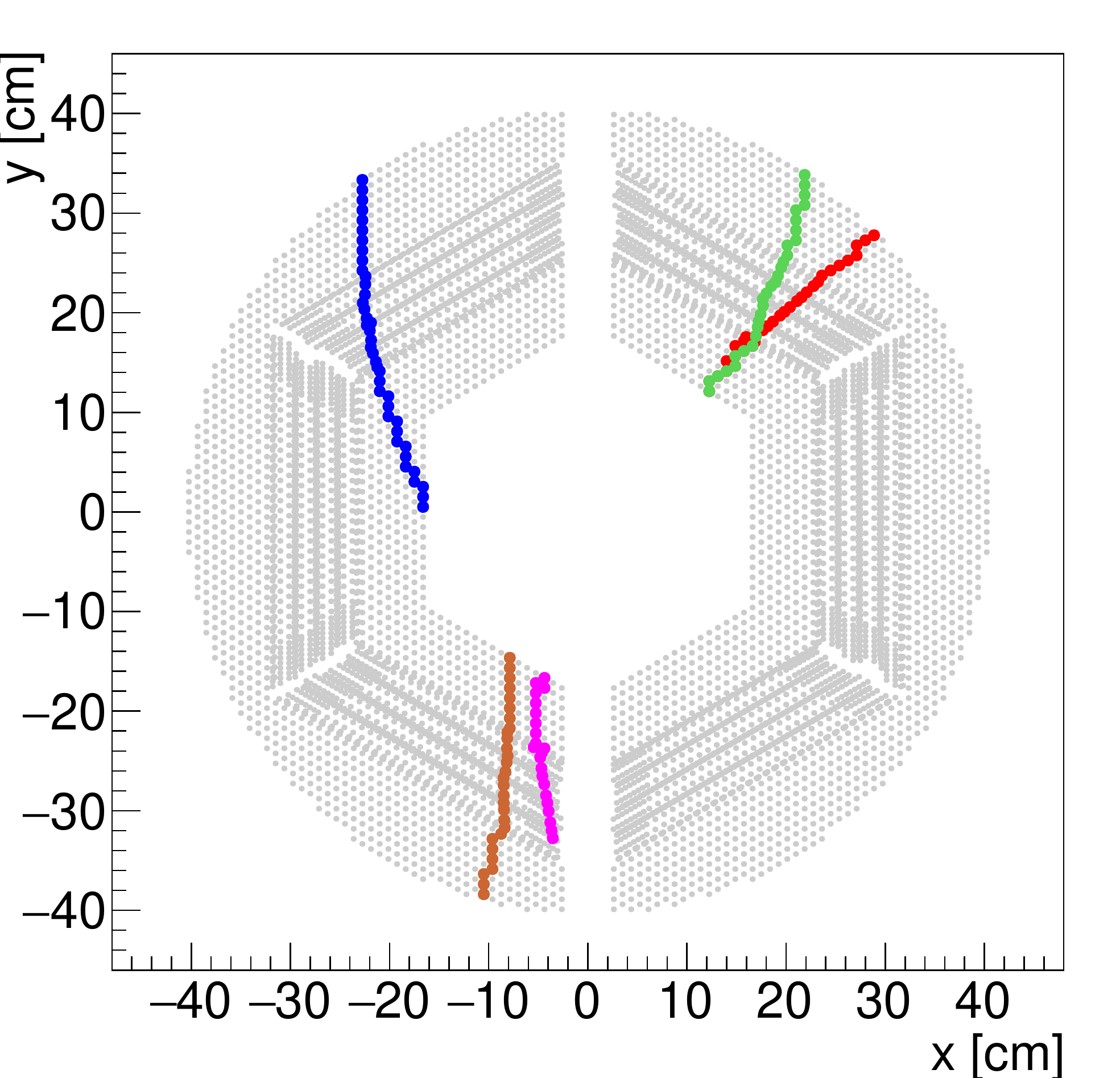}   \vspace{-0.1cm}
    \caption{An example of event reconstruction. Top row: the simulated paths of five particles shown in different colors (left) and the digitized tracks in the STT (right). Bottom row, left to right: the status of the reconstruction after the \textit{connect} phase (each different marker/color represents a different track candidate), the reconstruction after the \textit{fitting} phase (the black stars are set near tracks that will be merged in the third phase), and the final x-y reconstruction after the \textit{merging} phase. }
    \label{fig:conrec}
\end{figure*}

 \begin{figure*}[!htbp]
 \centering
    \includegraphics[width = 0.32\hsize]{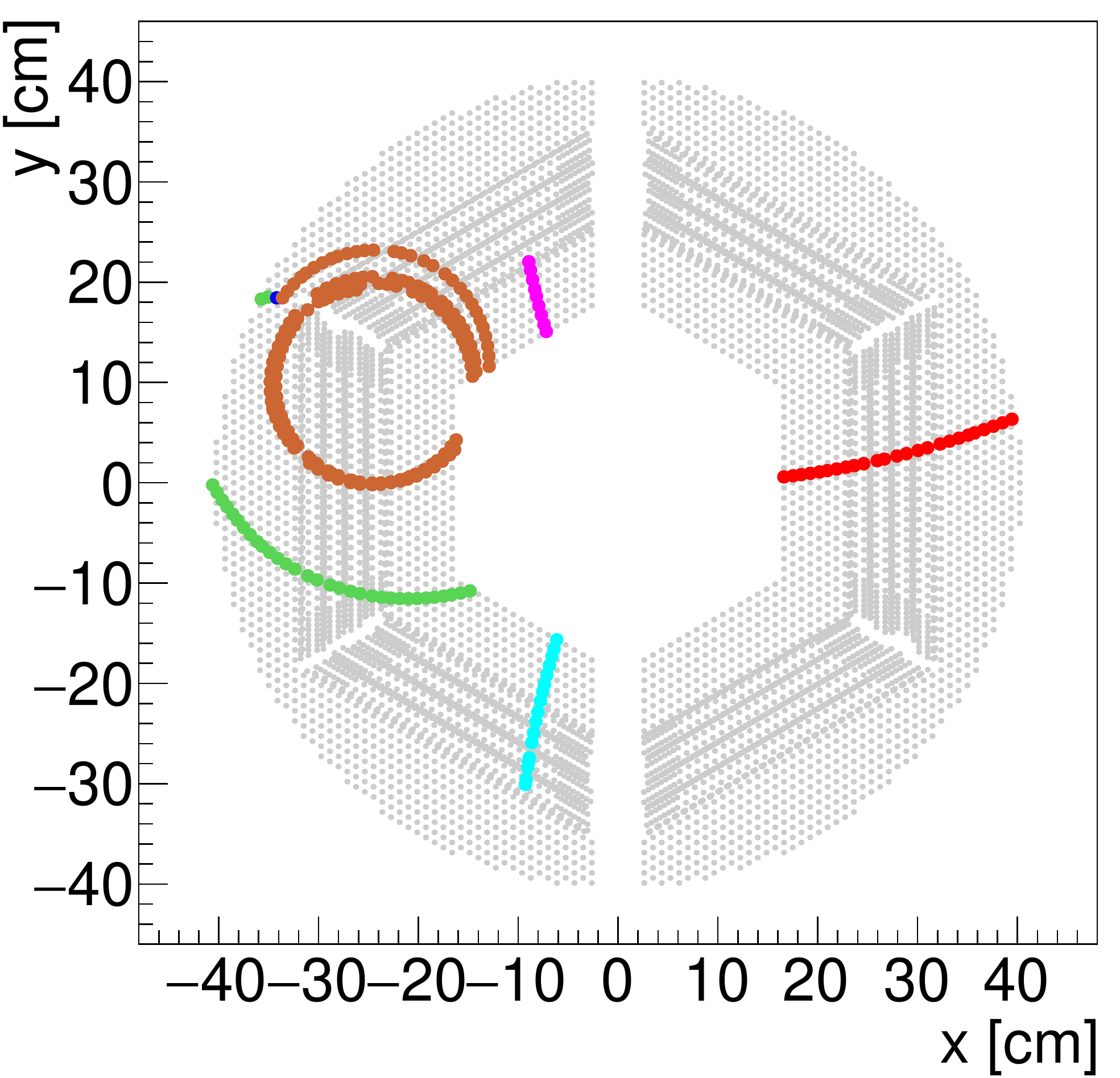} 
    \includegraphics[width = 0.32\hsize]{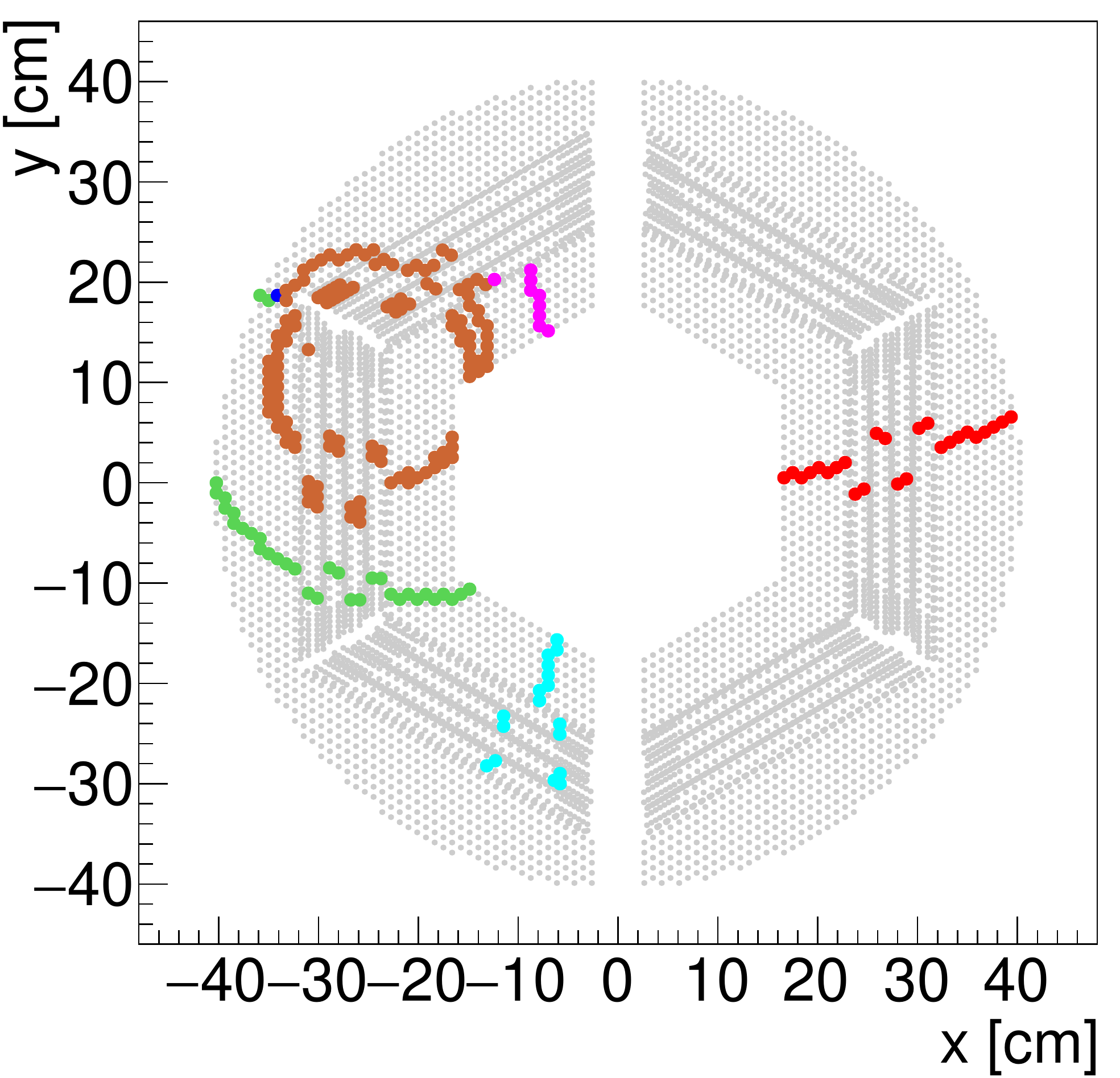}
    \includegraphics[width = 0.32\hsize]{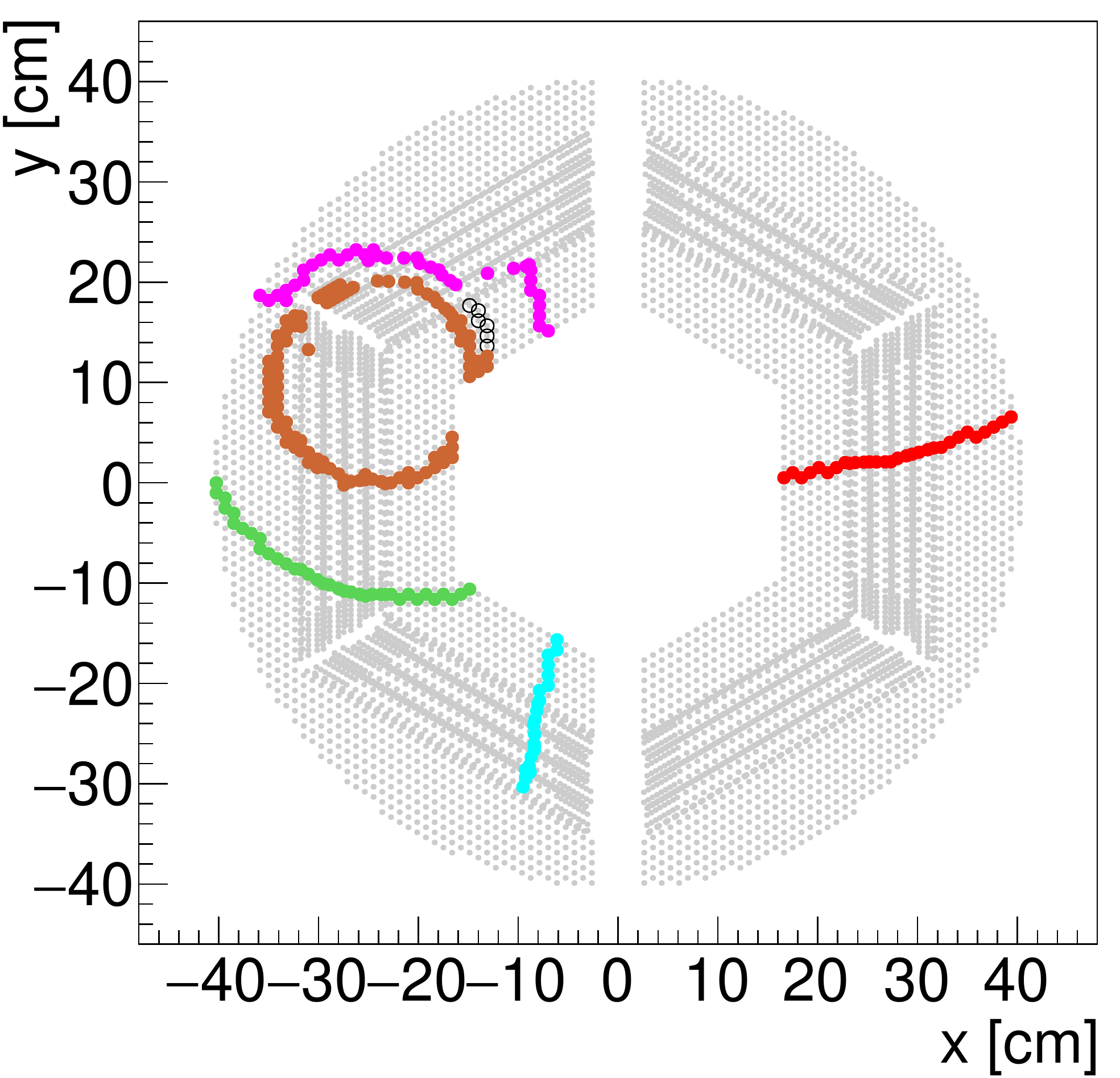}     

    \caption{Another example of track reconstruction for an event with five tracks. Left to right: the simulated particle trajectories, the digitized tracks, and the x-y reconstruction using the \textsc{lotf} algorithm.}
    \label{fig:evt1xy}
\end{figure*}    

\begin{figure*}[!htbp]
\centering
    \includegraphics[width = 0.34\textwidth]{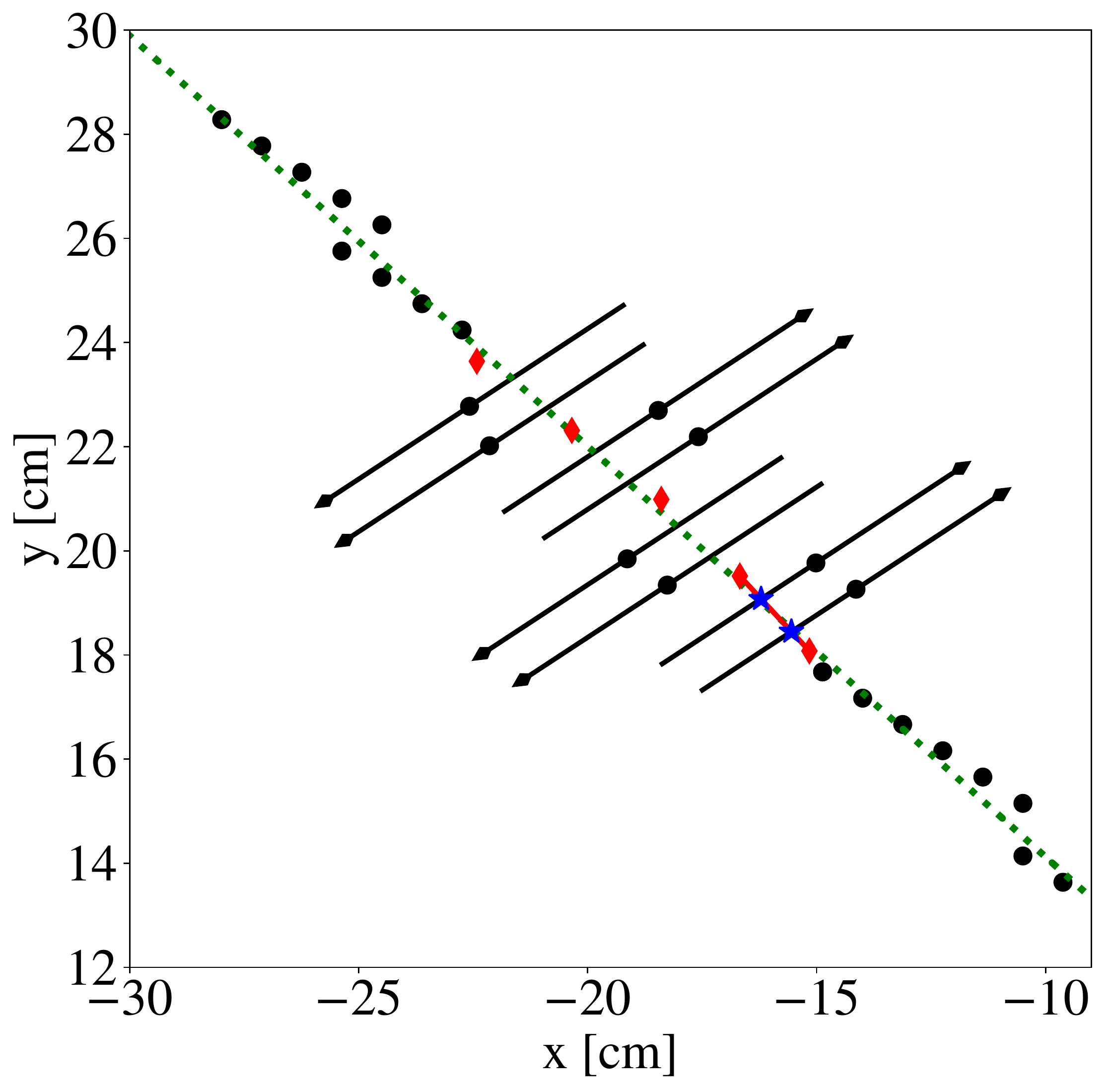} \qquad
    \includegraphics[width = 0.34\textwidth]{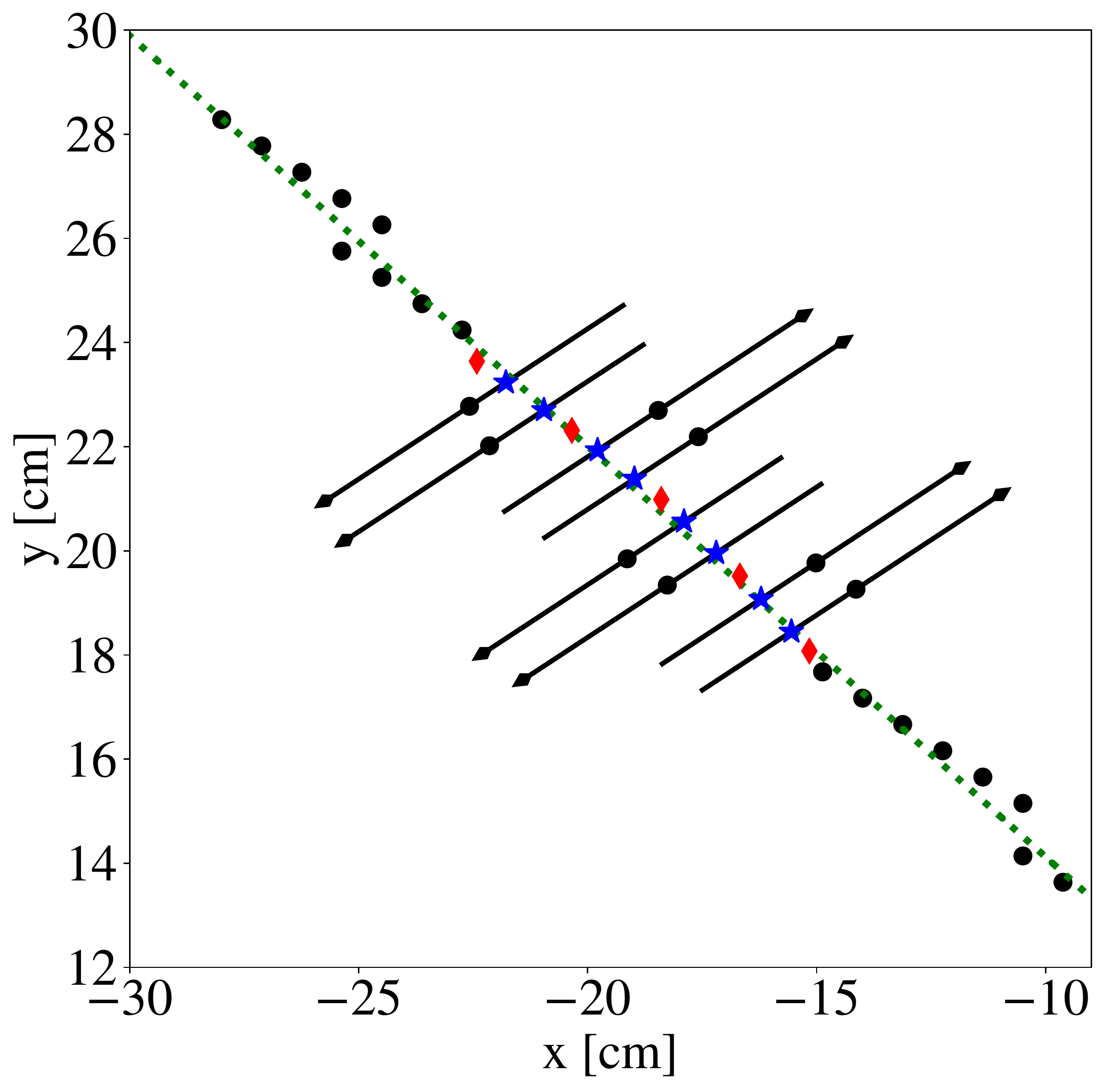} 
    \caption{An example illustrating how we use virtual nodes to refine the x-y hit coordinates along the skewed tubes. The simulated particle trajectory is shown with a green dotted line. The center-point coordinates of the activated tubes are indicated with black circles. The black arrows represent the skewed tubes, with the head pointing in different directions to denote the $+2.9$ and $-2.9$ degrees skew angles. Virtual nodes are shown with red diamonds. Left: we compute the intersection between the segment determined by two virtual nodes (red line) and the direction vector of the skewed tubes to refine the particle hit positions (blue stars) along the skewed tubes. Right: the final refined coordinates of the particle hits along all skewed tubes. For each skewed tube, determining the x-y hit coordinates precisely enables us to derive the z-information of the particle trajectory around these layers.}
    \label{fig:zskew}
\end{figure*}

  \begin{figure}[!htbp]
  \centering
       \includegraphics[width =0.8\hsize]{Images/evt0_simuzy.pdf}
       \includegraphics[width=0.8\hsize]{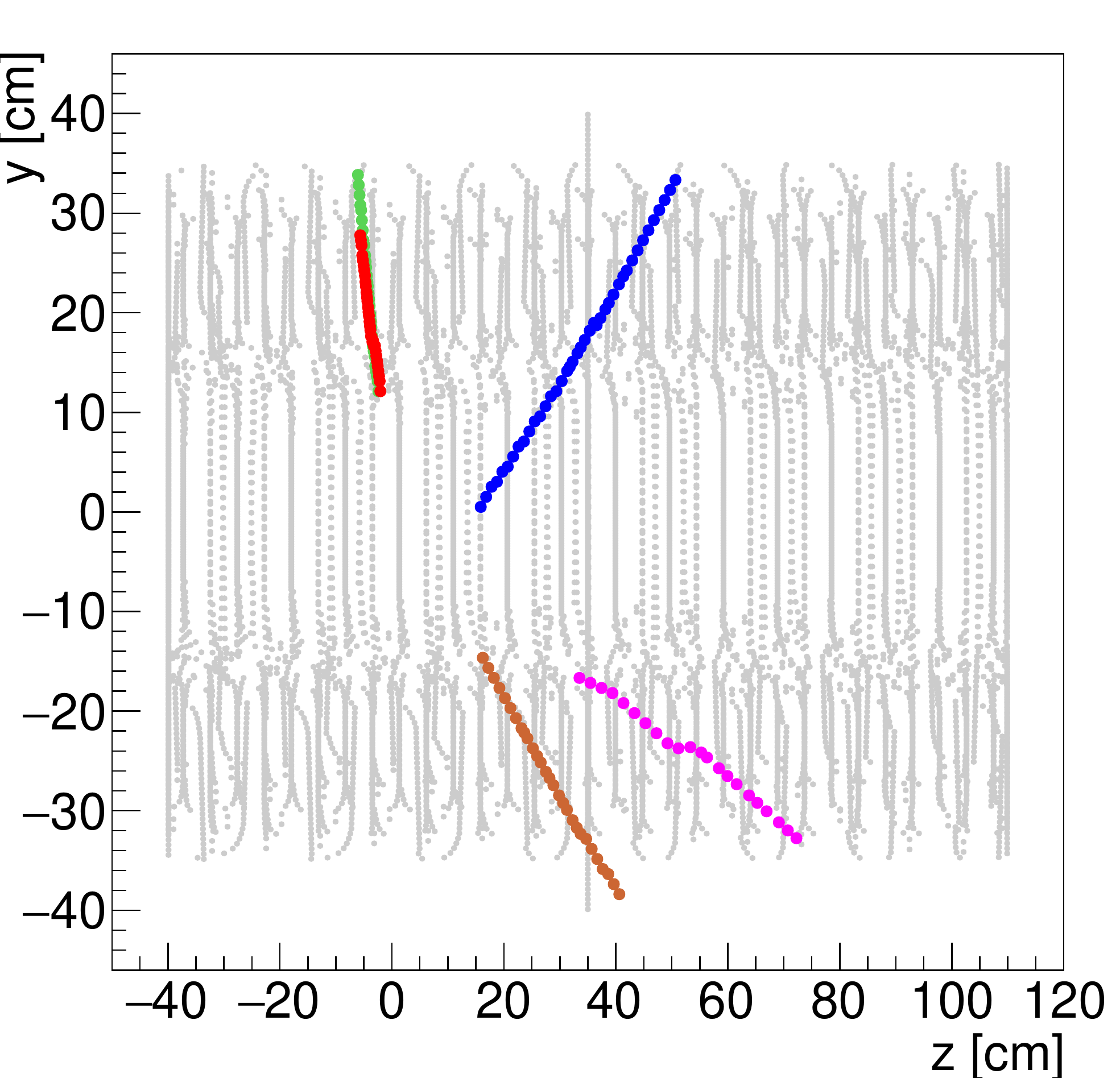}
    \caption{The result of the reconstruction of the z-information for the event A. Top: the simulated trajectories of the particles in the z-y plane. Bottom: the z-reconstruction obtained with the \textsc{lotf} algorithm using an interpolation system that fits a z-parametric line.}
      \label{fig:zcoord}
        \end{figure}
        
  \begin{figure}[!htbp]
\centering
    \includegraphics[width = .8\hsize]{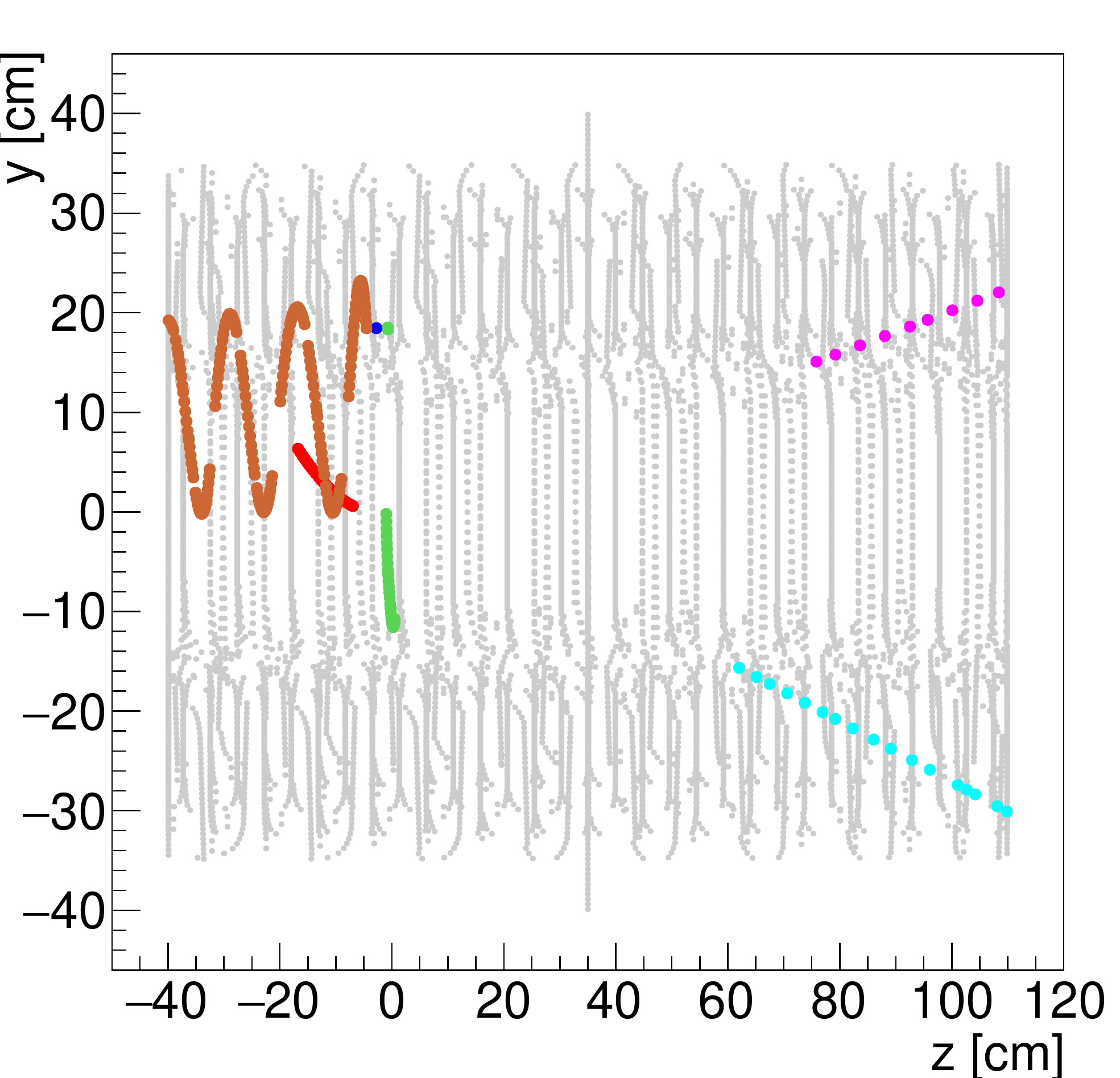}
    \includegraphics[width = .8\hsize]{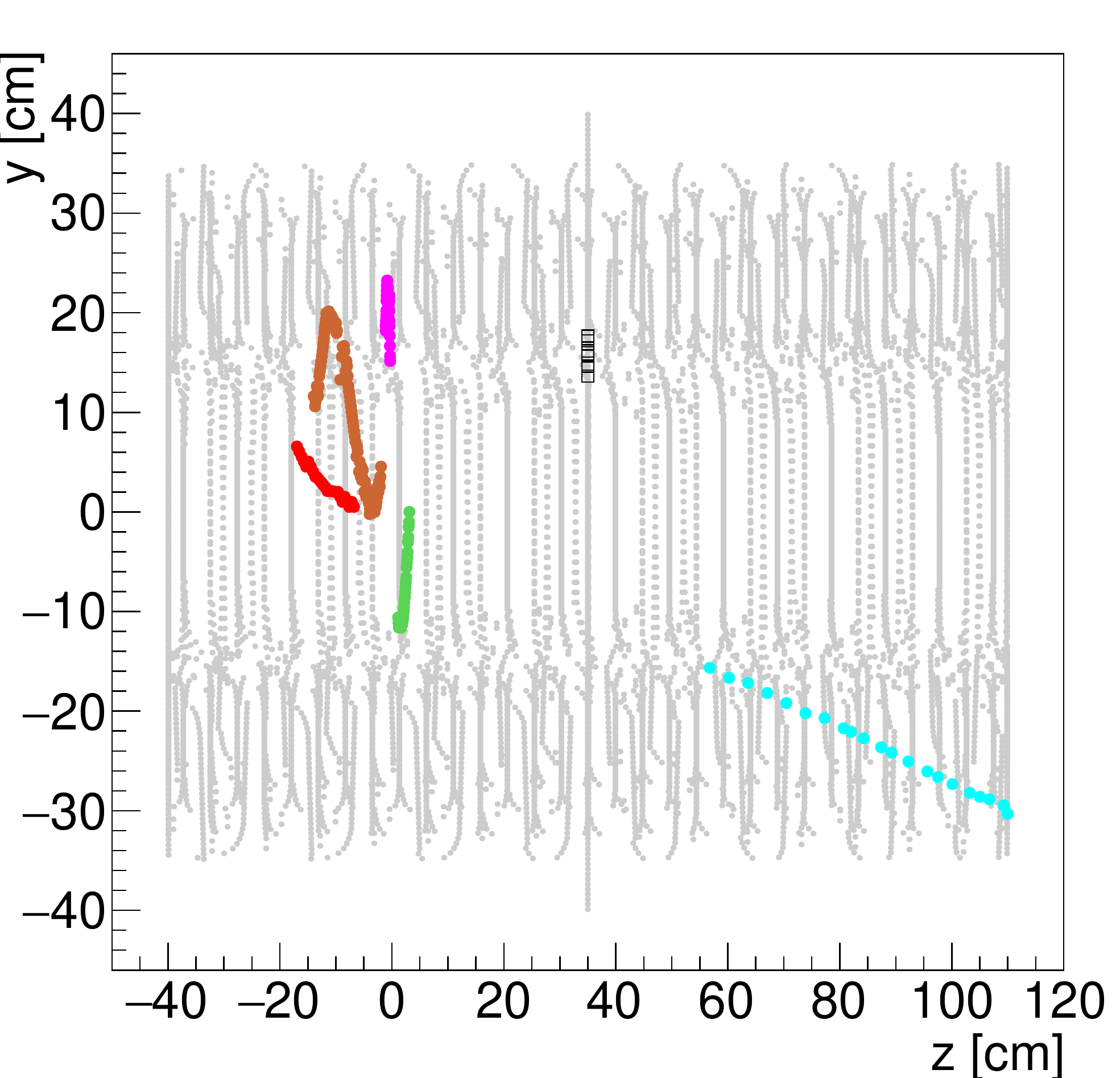} 
    \caption{Same as Figure~\ref{fig:zcoord} but for the event B. The z-information of the purple track is not correctly reconstructed because its x-y reconstruction includes nodes from two independent trajectories (see the discussion in the text). Additionally, we only recover a single loop from the brown trajectory because this particle passed through the same tubes several times. The \textsc{lotf} algorithm is not able to separate the distinct z-information from the individual loops that compose the full helical trajectory.}
    \label{fig:evt1yz}
  \end{figure}

In this section, we visualize two cases of track reconstruction in the x-y plane using the \textsc{lotf} algorithm. These examples are simply for illustration, a complete performance assessment is presented in Section~\ref{sect:perf}.

 Figure~\ref{fig:conrec} details the different steps of the track reconstruction algorithm for an event with five tracks (hereafter referred to as event A). Out of these five trajectories, one is isolated (upper left sector), two others are overlapping (upper right sector), and two are adjacent such that they have common neighbors (lower left sector). The bottom left panel presents the reconstruction status after the \textit{connect} phase. The bottom left panel shows that the blue trajectory is already purely reconstructed (we refer to a track as purely reconstructed when it contains all and only the hits from a single simulated track) after the \textit{connect} phase. The four tracks that are adjacent or overlapping are only partially reconstructed.
 
After the \textit{fitting} phase, the four incomplete tracks have been expanded such that they are either complete or flagged to be merged during the next phase (the black stars show the pair of tracks flagged for merging). 

Finally, at the end of the \textit{merging} phase, the brown and pink tracks are purely reconstructed, and the red and green reconstructions hold 94\% of all the hits from their respective simulated tracks. Additionally, the shape of the reconstructed tracks is consistent with the simulated trajectories shown in the top left panel. This is because, as further detailed in the next section, we re-determine each node's coordinates in order to correct for the discontinuities around the skewed layers.

%   \begin{figure}[!htbp]
%  \centering
% \includegraphics[width = 0.48\hsize]{Images/evt4_simuxy.pdf} 
%     \includegraphics[width = 0.48\hsize]{Images/evt4_read.pdf} \\
%     \includegraphics[width = 0.48\hsize]{Images/evt4_recxyTW.pdf} 
%     \includegraphics[width = 0.48\hsize]{Images/evt4_recxyBAB.pdf} 

%     \caption{An example of track reconstruction for an event with three particles. From top left to bottom right:  the simulated particle trajectories, the digitized tracks, the x-y reconstruction using our approach, and the reconstruction obtained with the \bab\ algorithm. In this event, there are hit gaps in the red digitized track such that no single continuous path connects all the nodes. Thanks to the procedure detailed in Section~\ref{sect:mergingg}, our algorithm provides a better reconstruction for this track.}
%     \label{fig:evt4xy}
% \end{figure}

Figure~\ref{fig:evt1xy} shows another event, hereafter referred to as event B, where a low-momentum secondary electron describes a circle in the STT volume (brown trajectory). The \textsc{lotf} algorithm provides an almost complete reconstruction of the trajectory of this particle: more than 76\% of the nodes in the simulated track have been recovered in a single reconstructed track, and only the outer part of the loop has been wrongly connected to another track (in purple).  The connection between the purple track and part of the brown trajectory is a consequence of our merging approach that considers tracks incomplete if they end in the middle layers of the STT. We plan to improve this issue in the future by refining the criterion that decides whether two tracks should be merged.
 
% We note that these low-momentum trajectories often correspond to secondary particles, generated due to interactions with detector material and not coming from the beam-target interaction point.  Reconstructing these tracks is not a priority in the context of the \panda\ experiment as they are typically contained within the STT volume and will lack information from other detectors which are relevant for the final event reconstruction. However, in case these tracks are overlapping with other trajectories, we must correctly identify them to efficiently discard them.  

\subsection{z coordinates reconstruction}
\label{sect:zrec}

%, however, the number of virtual nodes per track is limited by the number of times a particle transited through the skewed tubes layers. However, we can further benefit from these skewed tubes to recover additional The three-dimensional spatial extension of a skewed tube is strictly defined. Hence, determining the exact x-y coordinates of a particle hit on this tube directly provides the z coordinate as well. 
\begin{sloppypar}

By construction, the ability of the algorithm to accurately recover the z-information for each track depends on its x-y reconstruction performance. This is because, as detailed in Section~\ref{sect:zpro}, the z-reconstruction is based on the virtual nodes system, where each activated virtual node must be correctly assigned to the reconstructed track it belongs to. To achieve the z-reconstruction on the fly without using the tubes' drift time information, we use the well-defined coordinates of the virtual nodes to approximate the particle incident trajectory around the skewed tubes.

We implemented the procedure \textsc{CorrectSkewedXY}, the pseudo-code of which is given in Appendix~\ref{app:zrec}, to approximate the particle incident trajectory around the skewed layers. We describe its functioning here. Between two virtual nodes located on different layers, there are always two layers of skewed tubes. Hence, the local particle incident trajectory can be approximated using the line determined by two virtual nodes. The intersection between this line and the slope vector of the skewed tubes provides an estimate of the exact x-y position of the particle hit along the skewed tubes. An example detailing this procedure is presented in Figure~\ref{fig:zskew}. 
\end{sloppypar}

\begin{sloppypar}

Given the relatively small skew angle of these tubes, the typical uncertainty on the z coordinate determined in this way is significantly larger than the typical error on the x and y coordinates. For example, a $\pm$7.5~mm difference in the x-y estimates will lead to an uncertainty of $\pm$15~cm in the estimated z coordinate. However, as the number of z-hits recovered for each track increases, the error on the reconstructed z-momentum will decrease. We note that the determination of the z-information for a particle trajectory is only possible if the particle crossed at least two layers of tubes with different z-inclinations. 
\end{sloppypar}

\begin{sloppypar}
To obtain the z-information consistently for all the nodes in a track, we fit a parametric line in the z-y plane ($y = y_0 +y_1*t$ and $z = z_0 + z_1*t$ with $t$ the independent variable)\footnote{The choice of using the z-y plane instead of the z-x plane is arbitrary, and has no impact on the reconstruction results.} using all available z-coordinates from the virtual nodes and skewed tubes. e note that charged particles moving in a solenoidal magnetic field follow helical trajectories, hence, their projection onto the y-z plane repeats periodically rather than following a straight line. Nevertheless, for most particle tracks that are relevant in the context of the \panda\ experiment, the radius of the helix is much larger than the traversed STT dimensions. As a result, the curvature of the particle paths is not very noticeable when projected onto the y-z plane so the trajectories can be approximated by straight lines.
\end{sloppypar}

\begin{sloppypar}
We note that there exists some scatter in the reconstructed hit-coordinates, and these values are typically contained within a relatively small coordinate range ($\sim$10~cm for the y coordinates, see \textit{e.g.}, Figure~\ref{fig:virtex}). Additionally, we also lack the z-information in the inner part of the STT (between the beam-target interaction point and the STT inner radius of 16 cm). This means that additional constraints must be used to avoid getting unrealistic z-reconstruction.  In this work, we choose to add the point (x,y,z) = (0,0,0) to the list of values used for the z-reconstruction, but only for tracks that do not curl within the STT volume (\textit{i.e.}, tracks that have a clear inner-to-outer or outer-to-inner direction). This assumes that the origin of these tracks is close to the beam-target interaction point, which is a fair assumption for stable particles such as charged pions, protons, or antiprotons, that transverse the STT,  and short-lived particles, which decay almost instantly at the primary interaction point.  However, particles with a medium lifetime like ground state hyperons can travel several tenths of centimeters before they decay into stable particles. Here the assumption that the decay products come from the primary interaction point is not true. In these cases, our approach is not optimal and we plan to remove this assumption using the inner Micro Vertex Detector \citep[MVD,][]{pandamvd} that will allow us to improve the accuracy of the z-reconstruction for both the long-lived particles and their decay products. 
\end{sloppypar}

\begin{figure*}[!htbp]
    \centering
    \includegraphics[width = 0.9\hsize]{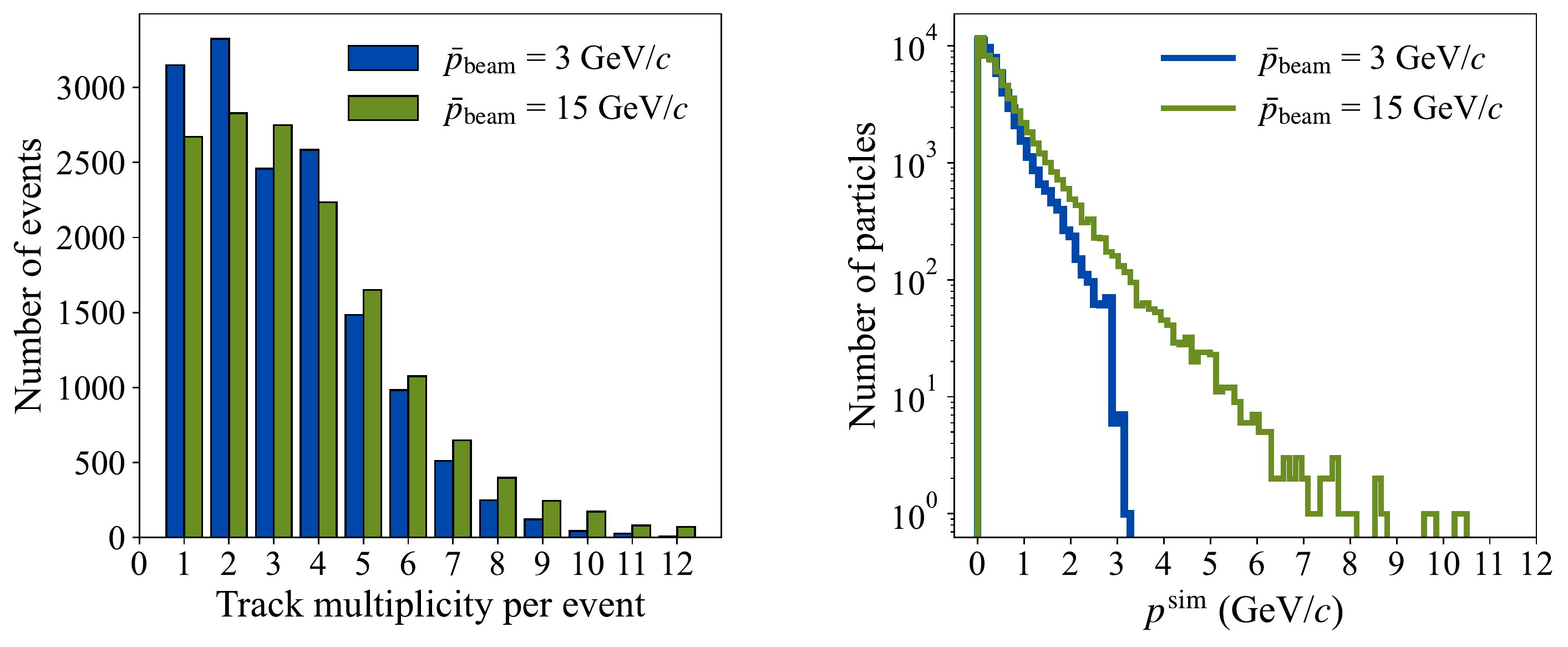}
    \caption{ Left: histogram representing the distribution of particle tracks per event in our experimental setup. A total of 30,000 events were simulated in the form of two times 15,000 events generated with an antiproton beam momentum ($\bar{p}_{\rm beam}$) of 3 and 15 GeV/$c$. Right: distribution of the particles' momentum for the 50,124 and 57,481 particles that reached the STT and have more than 5 STT hits in the simulations with a beam momentum of 3 GeV/$c$ and 15 GeV/$c$, respectively.}
    \label{fig:ntracks}
\end{figure*}

Figure~\ref{fig:zcoord} shows the z-reconstruction for event A.  The mean z-error (reconstructed minus simulated values) for the five tracks is 5.4 $\pm$ 2.5~cm which is relatively small compared to the typical z-extent of the simulated tracks. 

\begin{sloppypar}

Figure~\ref{fig:evt1yz} shows the z-reconstruction for event B. The green, red, and cyan tracks have a mean z-error of 2.4, 0.1, and 2.1~cm, respectively. The trajectory of the low-momentum secondary electron (brown) is only partially recovered in the z plane, yielding an average z-error of 16.4~cm. This particle passed through the same tubes several times, and \textsc{lotf} is not able to disentangle the individual loops that compose the full helical trajectory. Finally, the purple track has a mean z-error of 92~cm, emphasizing that the z information is poorly reconstructed. This is because the x-y reconstruction for this trajectory has merged portions of two independent simulated tracks (see Figure~\ref{fig:evt1xy}) such that the z-interpolation used coordinates from two separate trajectories. Consequently, the final z-reconstruction for this track is significantly different than the ground truth. This example emphasizes that the performance of the z-reconstruction crucially depends on the accuracy of the x-y reconstruction performed in the first place.
     
\end{sloppypar}

\section{Performance}
\label{sect:perf}

The purpose of \textsc{lotf} is to enable in-situ track reconstruction with the STT. To meet the requirements of the \panda\ physics experiment and have the highest chance of selecting events with rare or specific particle states, an optimal track reconstruction algorithm must have (1) large reconstruction efficiency (2) good reconstructed momentum resolution ($1-2\%$ \cite{Smyrski2012}), and (3), high reconstruction rates to match the expected 160 Mhits/s and 1,600 Mhits/s generated in the STT for the high-resolution and high-luminosity modes. In this work, we only assess the efficiency and reconstruction rate of the algorithm because both are most important for online track reconstructions. The reconstruction of the particles' momenta, which is performed after the track reconstruction and can be refined using the information from other trackers, will be investigated in a separate work.

We assess the performance of the \textsc{lotf} algorithm based on 30,000 events generated using the FTF background generator of the \textsc{PandaRoot} software\footnote{We used \textsc{PandaRoot} v12.0.1}. We simulated 15,000 of these events using an antiproton beam momentum ($\bar{p}_{\rm beam}$) of 3 GeV/$c$, and 15,000 events using a beam momentum of 15 GeV/$c$. We used two different beam-momentum values to generate events with different track multiplicities and particle momentum, $p$. The left panel of Figure~\ref{fig:ntracks} shows the frequency spectrum of the number of tracks per event. Most of the events have a track multiplicity between 1 and 4. On average, the  simulation using a 15 GeV/$c$ antiproton beam momentum has a larger track multiplicity per event than that of the  simulation with $\bar{p}_{\rm beam} =$ 3 GeV/$c$. The total number of STT tracks generated is 56,190 and 68,052 for the low and high beam momentum simulations, respectively. For the track analysis, we do not consider tracks that have fewer than 6 STT hits. The number of simulated tracks with at least 6 hits is 50,124 and 57,481 for the low and high beam momentum simulations, respectively.  

The right panel of Figure~\ref{fig:ntracks} shows the distribution of particle momentum simulated for each case. As expected, the distribution of particle momenta is more extended towards large $p$ when the antiproton beam momentum is 15 GeV/$c$. We note that the maximum $p$ obtained with the latter simulation is around 11 GeV/$c$, which is smaller than $\bar{p}_{\rm beam}$. This is because particles with larger momenta have very small scattering angles such that they do not cross the STT volume. Additionally, the momentum of each particle also depends on the number of particles that were formed during the interaction such that the distribution of individual $p$ is not immediately proportional to $\bar{p}_{\rm beam}$.  Overall, the majority of primary tracks originate from charged pions $\pi^+$ and $\pi^-$ but the type of simulated particles varies in the simulations such that we cover a large spectrum of event types. Of the 30,000 events, 21,376 ($\approx$71\%)  have overlapping tracks that are interesting for the performance assessment.

In Section~\ref{sect:phaseI}, we compare the performance of the \textsc{lotf} algorithm to the recent track reconstruction approach presented in \citet{babai20}, and to the current track reconstruction method, \textsc{BarrelTrackFinder}, implemented in the \textsc{PandaRoot} software. In Section~\ref{sect:phaseII}, we assess the algorithm's performance using data sets composed of 4 successive events to assess the impact of event-mixing. Finally, in Section~\ref{sect:time}, we scrutinize  the timing performance of \textsc{lotf}.

% [0.60922792 0.07572455 0.07871551 0.23633202 0.12930675 0.23171806]
% [0.41769547 0.0489117  0.22487481 0.30851802 0.16535277 0.29401557]
% [0.22156216 0.         0.41211267 0.36632517 0.19311803 0.26354872]

\begin{table*}[!htbp]
	\centering

	\begin{tabular}{cccccccccc} % four columns, alignment for each
	    \multicolumn{10}{c}{\large $\bar{p}_{\rm beam}$ = 3 GeV/$c$} \\
		\addlinespace[0.1cm]

		\hline
		Method & Simulated & Reconstructed & Found & F-Pure & F-Impure & P-Pure & P-Impure & Ghosts & Clones \\
		\hline	
% 		$\bar{p}_{\rm beam}$ = 3 GeV/$c$ &  \\ \\ 
		\multirow{2}{*}{\textsc{lotf}} & \multirow{6}{*}{\vspace{-0.2cm} 50,124}& \multirow{2}{*}{58,701} & {43,130} & 26,276 & 3,266 & 3,395 & 10,193& 5,577 & 9,994 \\
		& & & \textit{86.0\%} & 60.9\% & 7.6\% & 7.9\% & 23.6\% & 12.9\% & 23.1\% \\  %\textit{86.1\%}
		\addlinespace[0.1cm]
			\multirow{2}{*}{\textsc{Babai+20}} & &\multirow{2}{*}{58,868} & 40,338  & 16,849 & 1,973 & 9,071 & 12,445 & 6,670 & 11,860 \\
			& & & \textit{80.5\%} & 41.7\% & 4.9\% & 22.5\% & 30.9\% & 16.5\% & 29.4\% \\ %& \textit{19.5\%} 
			\addlinespace[0.1cm]

		\multirow{2}{*}{\textsc{BarrelTrackFinder}} & &\multirow{2}{*}{57,869} & 39,727  & 8,802 & 0 & 16,372 & 14,553 & 7,672 & 10,470 \\
		& & & \textit{79.3\%} & 22.1\% & 0.0\% & 41.2\% & 36.7\% & 19.3\% & 26.4\% 
		\\ \\%& \textit{20.7\%}
	\end{tabular}
% [0.61685007 0.09285537 0.07175658 0.21853798 0.13923134 0.17387887]
% [0.4130831  0.05690947 0.23668399 0.29332343 0.18170976 0.22257477]
% [0.16772319 0.         0.40269949 0.42957733 0.31982289 0.27137256]
	\begin{tabular}{cccccccccc} % four columns, alignment for each
	    \multicolumn{10}{c}{\large $\bar{p}_{\rm beam}$ = 15 GeV/$c$} \\
    		\addlinespace[0.1cm]

		\hline
		Method & Simulated & Reconstructed & Found &  F-Pure & F-Impure & P-Pure & P-Impure & Ghosts & Clones \\
		\hline	
% 		$\bar{p}_{\rm beam}$ = 3 GeV/$c$ &  \\ \\ 
			\multirow{2}{*}{\textsc{lotf}} & \multirow{6}{*}{\vspace{-0.2cm} 57,481}& \multirow{2}{*}{63,481} & 48,344 & 29,821 & 4,489 & 3,469 & 10,565 &  6,731 & 8,406   \\
			& & & \textit{84.1\%} & 61.7\% & 9.3\% & 7.1\% & 21.9\% & 13.9\% & 17.4\% \\  %& \textit{15.8\%} 
						\addlinespace[0.1cm]

			\multirow{2}{*}{\textsc{Babai+20}} & &\multirow{2}{*}{62,405} &  44,439 & 18,357 & 2,529 & 10,518 & 13,035 &  8,075 & 9,891   \\ %& 13,042
			& & & \textit{77.3\%} & 41.3\% & 5.7\% & 23.7\% & 29.3\% & 18.2\% & 22.2\% \\ %& \textit{22.7\%} 
						\addlinespace[0.1cm]

		\multirow{2}{*}{\textsc{BarrelTrackFinder}} & &\multirow{2}{*}{81,579} & 51,269 & 8,599 &  0 & 20,646 & 22,024 & 16,397  & 13,913 \\ % & 6,212
		& & & \textit{89.2\%}& 16.8\% & 0.0\% & 40.3\% & 42.9\% & 32.0\% & 27.1\%  \\ \\ %& \textit{10.8\%}
	\end{tabular}
		\caption{A summary of the performance of the three track reconstruction methods: \textsc{lotf} (this work), \textsc{Babai+20}, and \textsc{BarrelTrackFinder}. The top and bottom tables present the results of the track reconstruction for the low and high beam momentum simulations, respectively. The column ``Simulated" gives the number of simulated tracks that have at least 6 hits in the STT, and the column ``Reconstructed" shows the total number of reconstructed tracks for each approach. The column ``Found" displays the number of \textit{found} tracks, defined as the sum of all reconstructed tracks that are either F-Pure, F-Impure, P-Pure, or P-Impure (ranks 1 to 4, the description of these ranks is given in the text). The percentage shown in italic is the ratio of the number of \textit{found} tracks over the number of simulated tracks. The columns ``F-Pure" to ``Clones" detail the number of reconstructed tracks in each rank. All the percentages shown in these columns correspond to the ratio of the number of tracks in a given rank over the number of \textit{found} tracks (\textit{i.e.}, percentages from the columns ``F-Pure" to ``P-Impure" add up to 100).}
		\label{tab:res}
\end{table*}
\begin{figure*}[!htbp]
    \centering
    \includegraphics[width = 0.45\hsize]{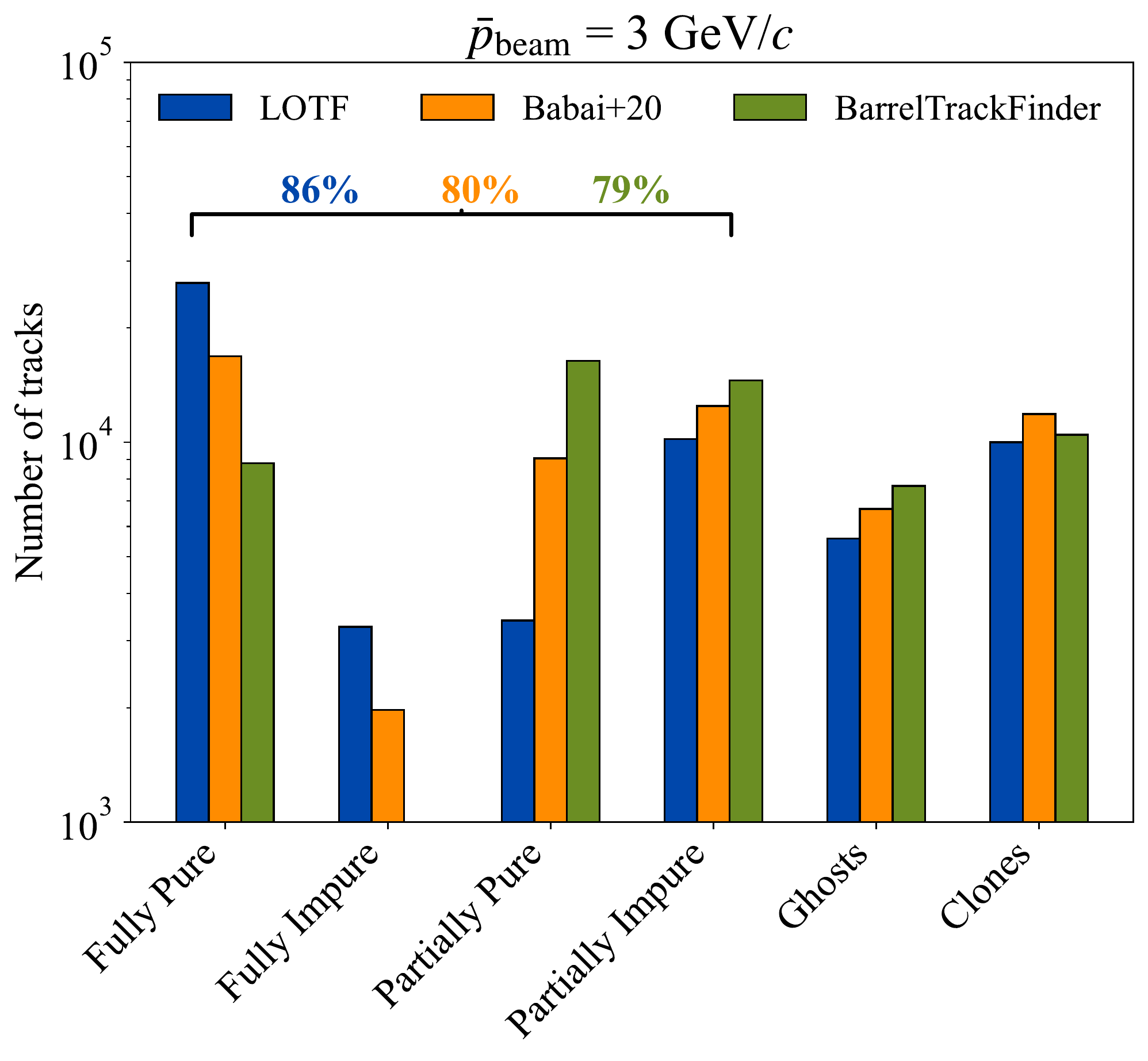}
    \includegraphics[width = 0.45\hsize]{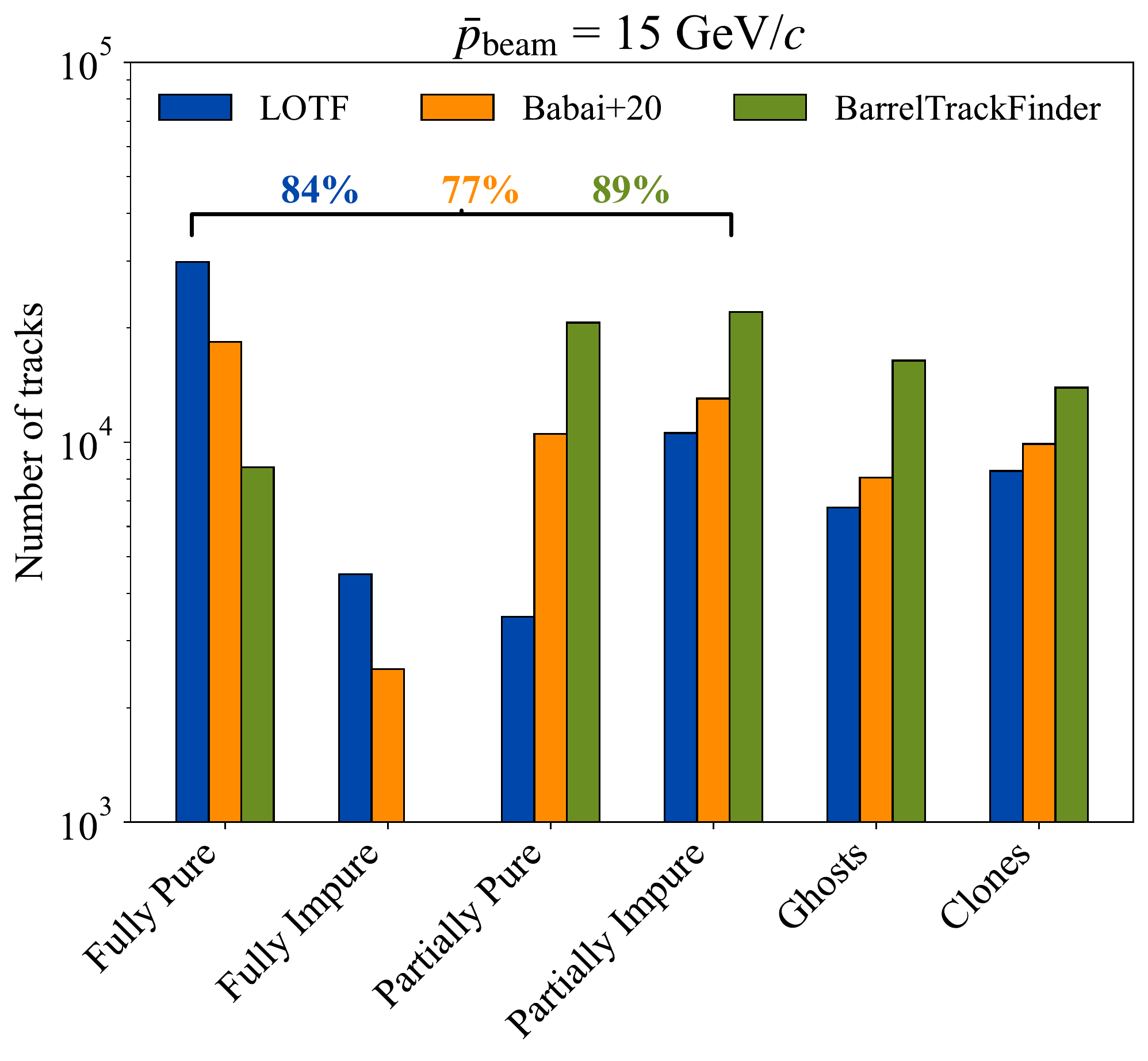}
    \caption{Left: histogram presenting the distribution of reconstructed tracks in each rank (defined in the text) for the 15,000 events with $\bar{p}_{\rm beam}$ = 3~GeV/$c$ (50,124 simulated tracks with more than 5 STT hits). Right: same but for the 15,000 events with $\bar{p}_{\rm beam}$ = 15~GeV/$c$ (57,481 simulated tracks with more than 5 STT hits). The percentages shown in each panel represent the ratio of the number of \textit{found} tracks (\textit{i.e.}, being either \textit{Fully Pure}, \textit{Fully Impure}, \textit{Partially Pure}, or \textit{Partially Impure}) over the number of simulated tracks for the three different methods. \textsc{lotf} has 86\% and 84\% of \textit{found} tracks in the simulations with $\bar{p}_{\rm beam}$~=~3~GeV/$c$ and $\bar{p}_{\rm beam}$~=~15~GeV/$c$, respectively. Additionally, it has the largest number of \textit{Fully Pure} tracks (perfectly reconstructed tracks) and the lowest number of \textit{Ghosts} and \textit{Clones} (incorrect reconstructions) in both cases.} 
    \label{fig:qa}
\end{figure*}

\subsection{Track reconstruction performance}
\label{sect:phaseI}

\begin{sloppypar}
Here we focus on assessing the performance of \textsc{loft} in assigning the correct nodes to the correct track (Section~\ref{sect:compari}), while simultaneously quantifying the corresponding average z-reconstruction error (Section~\ref{sec:zrecperf}).
\end{sloppypar}

\subsubsection{Comparison with other methods}
\label{sect:compari}

\begin{sloppypar}  In the context of cylindrical straw tube detectors, several track reconstruction approaches have been developed over the past years (e.g. \cite{Schumann:139096, herten2015gpu, Regina:phd, Andersson2021AGA, Alicke:2021omm, Liang2017, Karabowicz, babai20}). In this work, we compare \textsc{lotf} to two of these approaches: the track reconstruction technique from \citet{babai20} (referred to as \textsc{Babai+20}) and the \textsc{BarrelTrackFinder} algorithm  \cite{Karabowicz} implemented in \textsc{PandaRoot} \cite{Spataro2011}. 

The \textsc{BarrelTrackFinder} algorithm builds upon conformal mapping \cite{Hoppner:2009af}; circles in the x-y plane are transformed into straight lines and hits are merged based on their connections in this new plane. The \textsc{BarrelTrackFinder} can perform a global track reconstruction including information from several PANDA detectors. For a fair comparison, we only use it locally by including the STT information. 

\textsc{Babai+20} performs charged-particle trajectory reconstruction using attribute-space-connected morphological filters \cite{wilkinson07:_attrib_space_connec_connec_filter}; the original data set is transformed into an orientation-based space where hits are grouped into tracks depending on which orientation subspace they belong to.  The \textsc{Babai+20} approach is different from the local linear fitting procedure used in \textsc{lotf}, yet, \textsc{lotf} builds upon several techniques introduced by \textsc{Babai+20} such as the construction of a connected graph representing the STT geometrical information and the use of virtual nodes. Hence, the comparison to \bab\ is relevant to explore the performance improvements enabled by \textsc{lotf}.   

To assess the track reconstruction performance, we use the metric extended and described in \cite{Regina:phd} and implemented in the \textsc{PandaRoot} software: the reconstructed tracks are assigned a \textit{rank} which characterizes the quality of the reconstruction. These ranks are defined as follows:

\begin{enumerate}[labelsep=0ex,align=left]
    \item[Rank 1. ] \textit{Fully Pure} track: the reconstructed track includes only the hits belonging to a single simulated track, and all hits from the latter have been found. In other words, the reconstructed track perfectly matches a given simulated track.
    \item[Rank 2. ] \textit{Fully Impure} track: the reconstructed track holds all the hits from a simulated track, but also includes some impurities, \textit{i.e.}, hits belonging to other simulated tracks. The fraction of \textit{impure} hits must not be larger than 30\%.
    \item[Rank 3. ] \textit{Partially Pure} track: the reconstructed track holds at least 70\% of all the hits from a simulated track and only includes hits from that track (\textit{i.e.}, no impurities allowed).
    \item[Rank 4. ] \textit{Partially Impure} track: at least 70\% of all the hits in the reconstructed track originate from a single simulated track. A Rank 4 track might include impurities.
    \item[Rank 5. ] \textit{Ghost} track: less than 70\% of the hits in the reconstructed track originate from the same simulated track. 
    \item[Rank 6. ] \textit{Clone} track: if several reconstructed tracks match with the same simulated track, only the reconstructed track with the largest intersection set is defined as the \textit{true} reconstructed track, the other reconstructions are clones. 
\end{enumerate}

All the reconstructed tracks can only belong to one of these categories. Rank 1 is the best achievable rank, the total number of tracks having a rank comprised between 1 and 4 are regarded as \textit{found} tracks while the number of \textit{Clones} and \textit{Ghosts} tracks (ranks 5 and 6) are regarded as incorrect reconstructions.  An optimal track reconstruction algorithm should have the largest number of \textit{found} tracks, the lowest number of \textit{Clones} and \textit{Ghosts}, and the largest number of \textit{Fully Pure} tracks.

Using this metric, we compare the three algorithms in reconstructing the 50,124 and 57,481  tracks simulated with the FTF generator having at least 6 hits in the STT in the low and high beam momentum simulations, respectively. Table~\ref{tab:res} summarizes the number and percentages of reconstructed tracks in each category for the three methods. The percentage of \textit{found} track is derived with respect to the total number of simulated tracks, and the percentages of tracks in each rank are derived with respect to the number of \textit{found} tracks. Hence, for tracks that are either \textit{Fully Pure}, \textit{Fully Impure}, \textit{Partially Pure}, or \textit{Partially Impure}, these percentages represent the fraction of \textit{found} tracks in each category. For the \textit{Clones} and \textit{Ghosts}, these percentages represent the fraction of incorrect tracks reconstructed in addition to the \textit{found} tracks.

Figure~\ref{fig:qa} shows these results in the form of histograms for the two simulation data sets.  \textsc{lotf} has the largest percentage of \textit{found} tracks in the simulation with $\bar{p}_{\rm beam}$~=~3~GeV/$c$ (86\%), and the second largest in the simulation with $\bar{p}_{\rm beam}$~=~15~GeV/$c$ (84\%, while the \textsc{BarrelTrackFinder} performs better with 89\%). Interestingly, for the \textsc{BarrelTrackFinder} approach, the fraction of \textit{found tracks} increases in the simulation with $\bar{p}_{\rm beam}$~=~15~GeV/$c$ compared to the simulation with  $\bar{p}_{\rm beam}$~=~3~GeV/$c$ (89\% against 79\%), while \textsc{lotf} and \textsc{Babai+20} have the opposite behavior (a drop of 2 to 3\% in the number of \textit{found} tracks for both). We note that \textsc{BarrelTrackFinder} has a factor of 1.5 to 2.5 times more \textit{Ghosts} and \textit{Clones} in the simulation with $\bar{p}_{\rm beam}$~=~15~GeV/$c$ than \textsc{lotf} and \textsc{Babai+20}, suggesting that its better performance comes at a cost of having a larger number of incorrect reconstructions.

Figure~\ref{fig:qa} also highlights that \textsc{lotf} holds the largest number of \textit{Fully Pure} tracks, a number 1.6 to 4 times larger for \textsc{lotf} than for \textsc{Babai+20} and \textsc{BarrelTrackFinder}, respectively. The fraction of these tracks in the low (high) beam momentum simulation case is 60.9\% (61.7\%) for \textsc{lotf}, 41.7\% (41.3\%) for \textsc{Babai+20}, and 22.1\%  (16.8\%) for \textsc{BarrelTrackFinder}. Additionally, \textsc{lotf} has the lowest amount of \textit{Ghosts} and \textit{Clones} of all three methods for both data sets.

Finally, to process the 30,000 events using an Intel i9-11900H CPU at 2.50~GHz, \textsc{lotf} took 44.3~s, \textsc{BarrelTrackFinder} 590.1~s, and \textsc{Babai+20} 41,301.4~s. An in-depth analysis of the \textsc{lotf} processing time as a function of the different reconstruction phases is presented in Section~\ref{sect:time}.  
% Overall, in both the low and high beam momentum simulations, \textsc{lotf} has (1) a large fraction of \textit{found} tracks (86\% and 84\% for the simulations with $\bar{p}_{\rm beam}$~=~3~GeV/$c$ and $\bar{p}_{\rm beam}$~=~15~GeV/$c$, respectively), (2) the largest number of \textit{Fully Pure} tracks (rank 1, perfect reconstructions), and (3) the lowest amount of \textit{Ghosts} and \textit{Clones} (incorrect reconstructions).  Hence, \textsc{lotf} has both the best track reconstruction performance and computational time which makes it a promising method for enabling an efficient in-situ event selection in \panda. 

% Finally, as shown in the last column of Table~\ref{tab:res}, \textsc{lotf} has the lowest percentage  of \textit{not found} tracks (\textit{i.e.}, simulated tracks that have not been matched to any reconstructed tracks) for the simulations using a 3~GeV/$c$ beam momentum (13.9\%, compared to 19.5\% and 20.7\% for \textsc{Babai+20} and \textsc{BarrelTrackFinder}, respectively). In the simulation using a 15~GeV/$c$ antiproton beam momentum, \textsc{BarrelTrackFinder} performs better with only 10.8\% of \textit{not found} tracks being, against 15.8\% for \textsc{lotf} and 22.7\% for \textsc{Babai+20}. Yet, as noted above, \textsc{BarrelTrackFinder} has a significantly larger amount of incorrect reconstructions in the latter case.\\

\end{sloppypar}

\subsubsection{z-reconstruction performance}
\label{sec:zrecperf}

In this section, we investigate the z-reconstruction performance of \textsc{lotf}.  We define the ``z-error"  as z$_{\rm rec} -$ z$_{\rm sim}$, where z$_{\rm rec}$ and z$_{\rm sim}$ are the reconstructed and simulated z-coordinates of the particle hits, respectively. To compute these values, we match, for each event,  each simulated track with the reconstructed track that has the largest intersection set (larger number of matching hits). We then compute the z-errors for each node in the intersection set. \textsc{lotf} reconstructs the z-information using virtual nodes, a technique introduced in \citet{babai20}. Hence, we choose to compare the z-reconstruction performance of both approaches. We do not include \textsc{BarrelTrackFinder} in this comparison as there is no simple way to extract the reconstructed z coordinates for the latter method. We leave an in-depth analysis of the particles' longitudinal momentum reconstruction, including a comparison with recent promising approaches (\textit{e.g.}, the \textsc{PzFinder}  \cite{Andersson2021AGA}), for future work.

\begin{figure}[!htbp]
    \centering
        \includegraphics[width = 0.95\hsize]{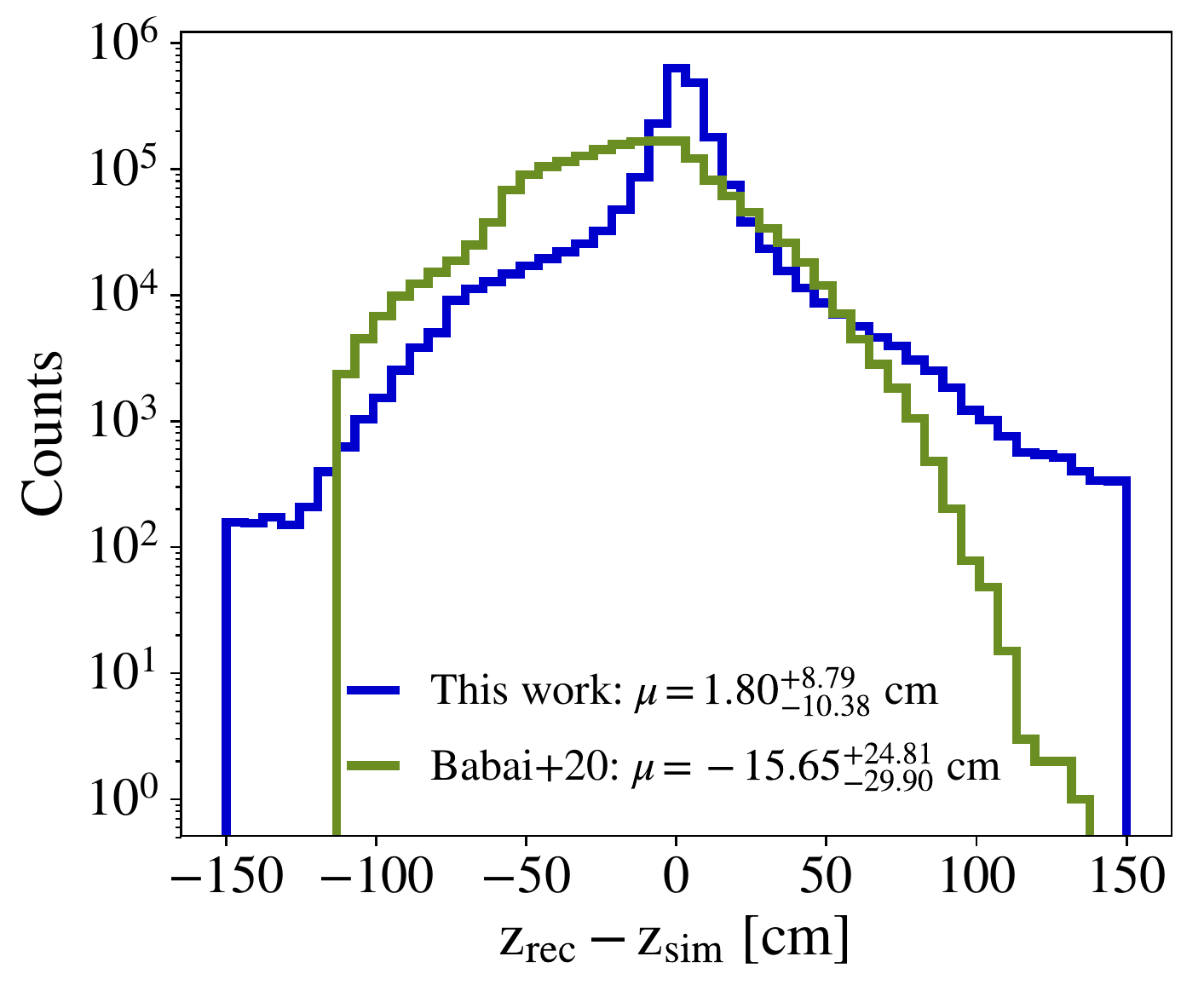}
    \caption{ The distribution of the z-errors defined as the difference between the z-values of the reconstructed tracks (z$_{\rm rec}$) and the z-values in the simulation (z$_{\rm sim}$). The blue curve shows the distribution for the \textsc{lotf} algorithm and the green curve shows that of \bab. The median and associated 16$^{\rm th}$ and 84$^{\rm th}$ percentile errors is $1.80^{+9.79}_{-10.38}$~cm for \textsc{lotf}  and $-15.65^{24.81}_{29.90}$~cm for  \bab. \textsc{lotf} improves the median z-error by a factor of 8.7 compared to \bab. }
    \label{fig:babz}
\end{figure}

\begin{sloppypar}

 Figure~\ref{fig:babz} compares the distribution of the z-errors for \bab\ and \textsc{lotf} for all the reconstructed tracks in the 30,000 events. As noted in Section~\ref{sect:zrec}, the z-information can only be reconstructed if the particle crossed at least two skewed layers in the STT. Among the 50,124 and 57,481 tracks from the low and high beam momentum simulations considered in this experiment, 43,583 and 48,712 fulfill this requirement, respectively, and are used for assessing the z-reconstruction performance.
The \bab\ yields a median z-error of $-15.65^{24.81}_{29.90}$~cm, a value 8.7 times larger than the $1.80^{+9.79}_{-10.38}$~cm median z-error obtained with \textsc{lotf}. Additionally, 68.3\% of the reconstructed z-values  are within $\pm$10~cm of the simulated values for \textsc{lotf}, while this percentage is only 26.4\% for \bab. 

The wings of the distribution of the z-errors with \textsc{lotf} are wider than for \bab.  The number of tracks that have an average z-reconstruction error of more than 110~cm is 77 for \textsc{lotf}, but only one for \bab.  The absence of very large under- or over-estimation of the z-coordinates for the \bab\ can be explained by the different approaches used by the authors to derive the virtual node coordinates. As mentioned in Section~\ref{sect:zrec}, the virtual node z-coordinates lie in a limited range, typically between 0 to $\sim$60~cm, such that the z-error distribution is typically contained between -120 and +120~cm. While the number of cases where \textsc{lotf} gives very large under- and over-estimation of the z-coordinates is low ($< 0.1\%$ of all tracks), these cases could be prohibitive during actual experiments because they would lead to wrong estimates of the particles' longitudinal momentum component. Nevertheless, in practice, including information from the inner detectors such as the MVD would enable significant improvements in the z-reconstruction performance, and this aspect will be explored in future work.
 \end{sloppypar}

 Overall, these results highlight that \textsc{lotf} recovers the z-coordinates with a relatively small error compared to the overall STT longitudinal dimension ($\sim$150~cm). This aspect is promising for extracting an accurate estimate of the particles' longitudinal momentum component.

\begin{figure*}[!htbp]
\centering
    \includegraphics[width = 0.32\hsize]{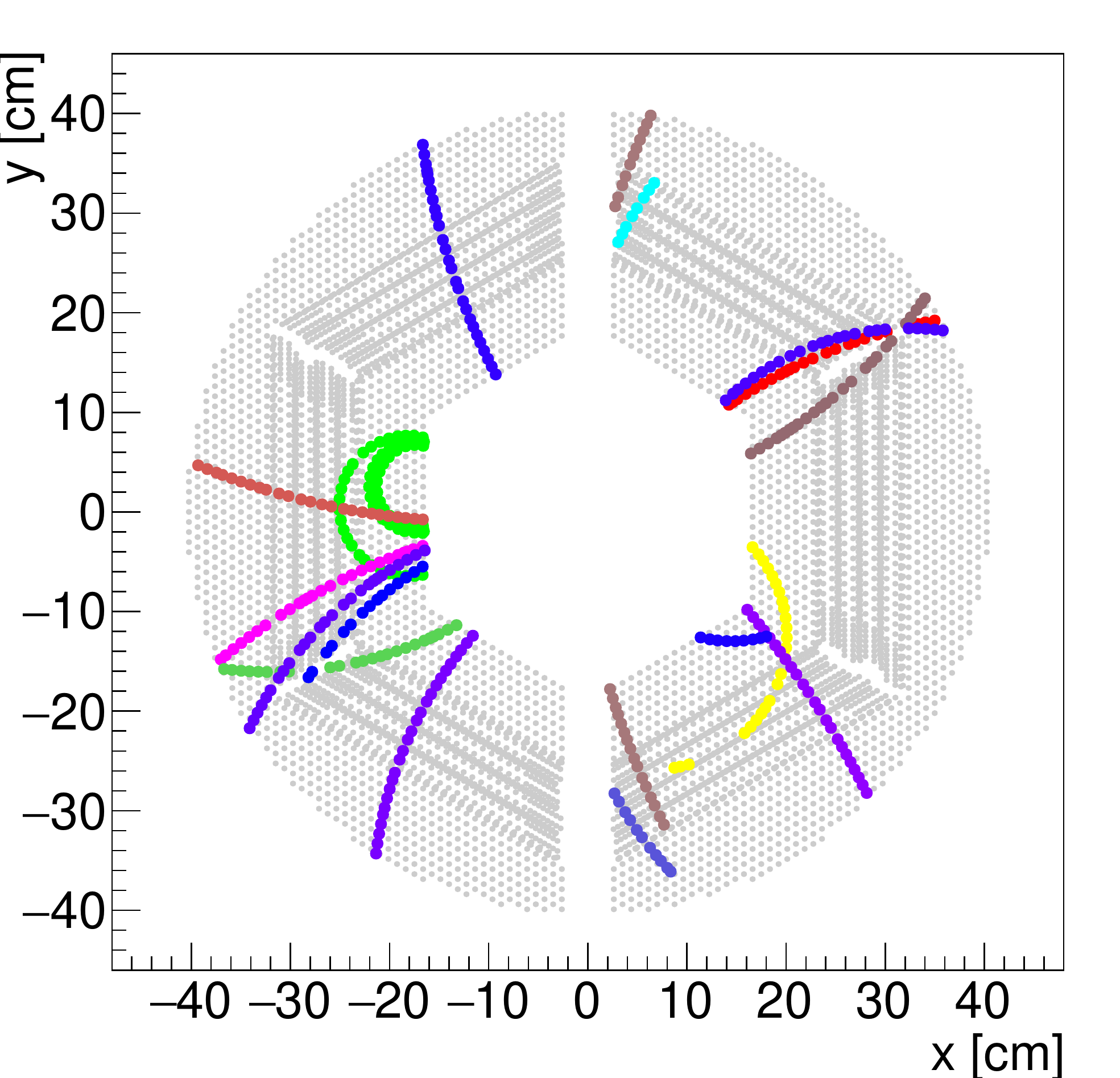}
    \includegraphics[width = 0.32\hsize]{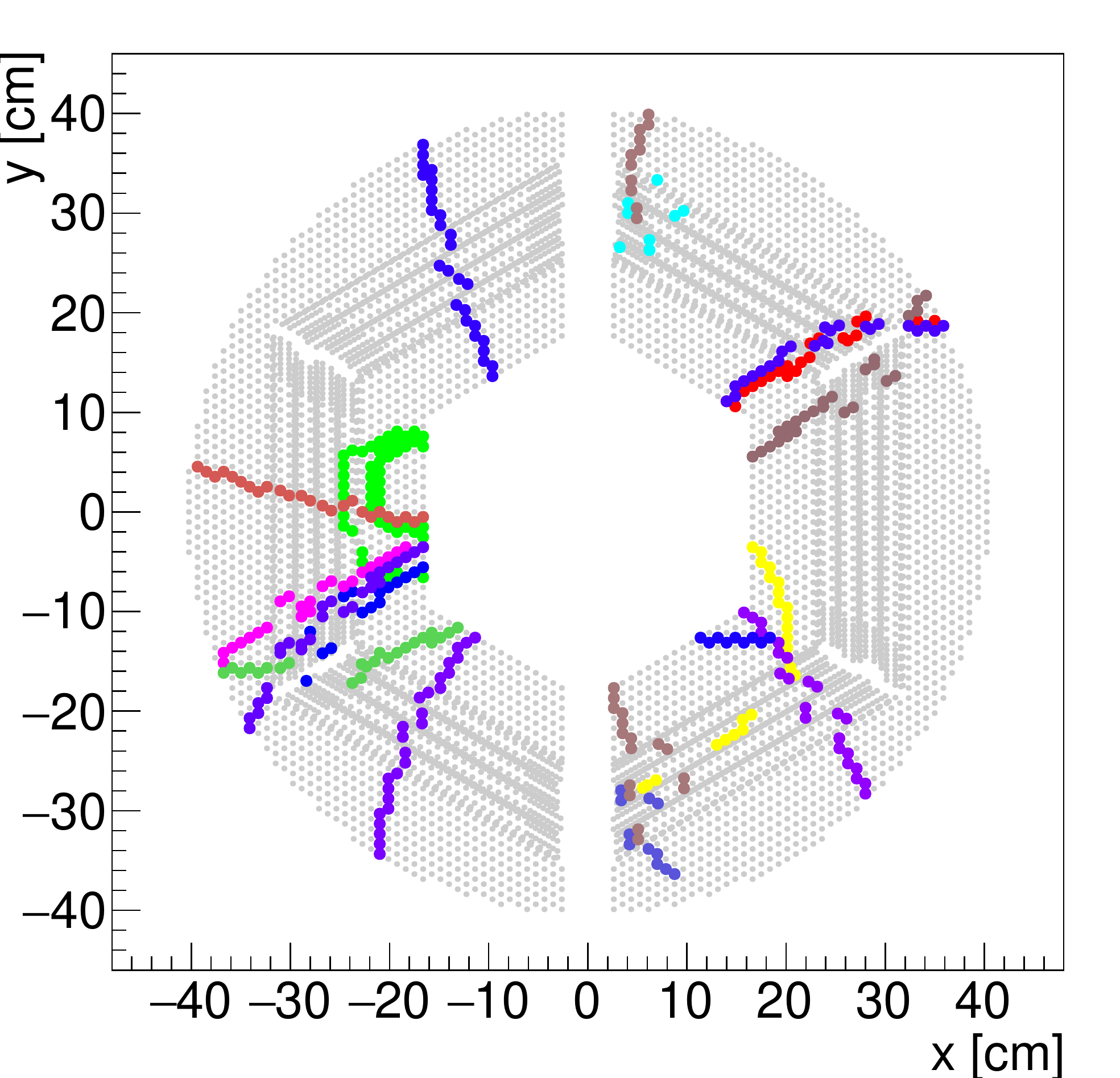}
    \includegraphics[width = 0.32\hsize]{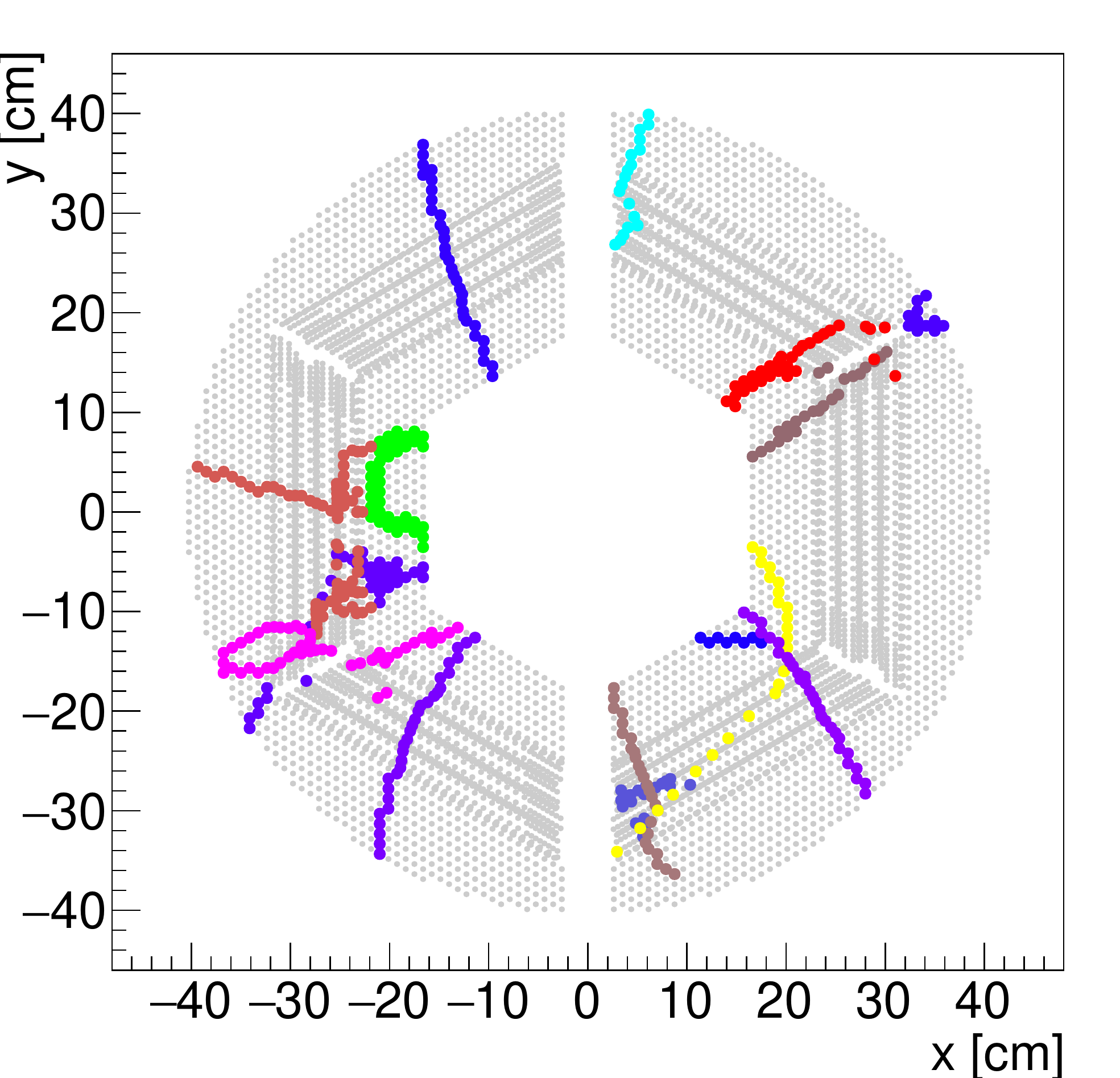}
    \caption{An example of data set used to test our algorithm in Section~\ref{sect:phaseII}. We concatenated 4 successive events to simulate the effects of event-mixing. Left: The simulated trajectories. Middle: The corresponding hits in the STT. Right: The reconstruction using our algorithm.}
    \label{fig:evtcomp}
\end{figure*}

\begin{figure}[!htbp]
    \centering
    \includegraphics[width = 0.95\hsize]{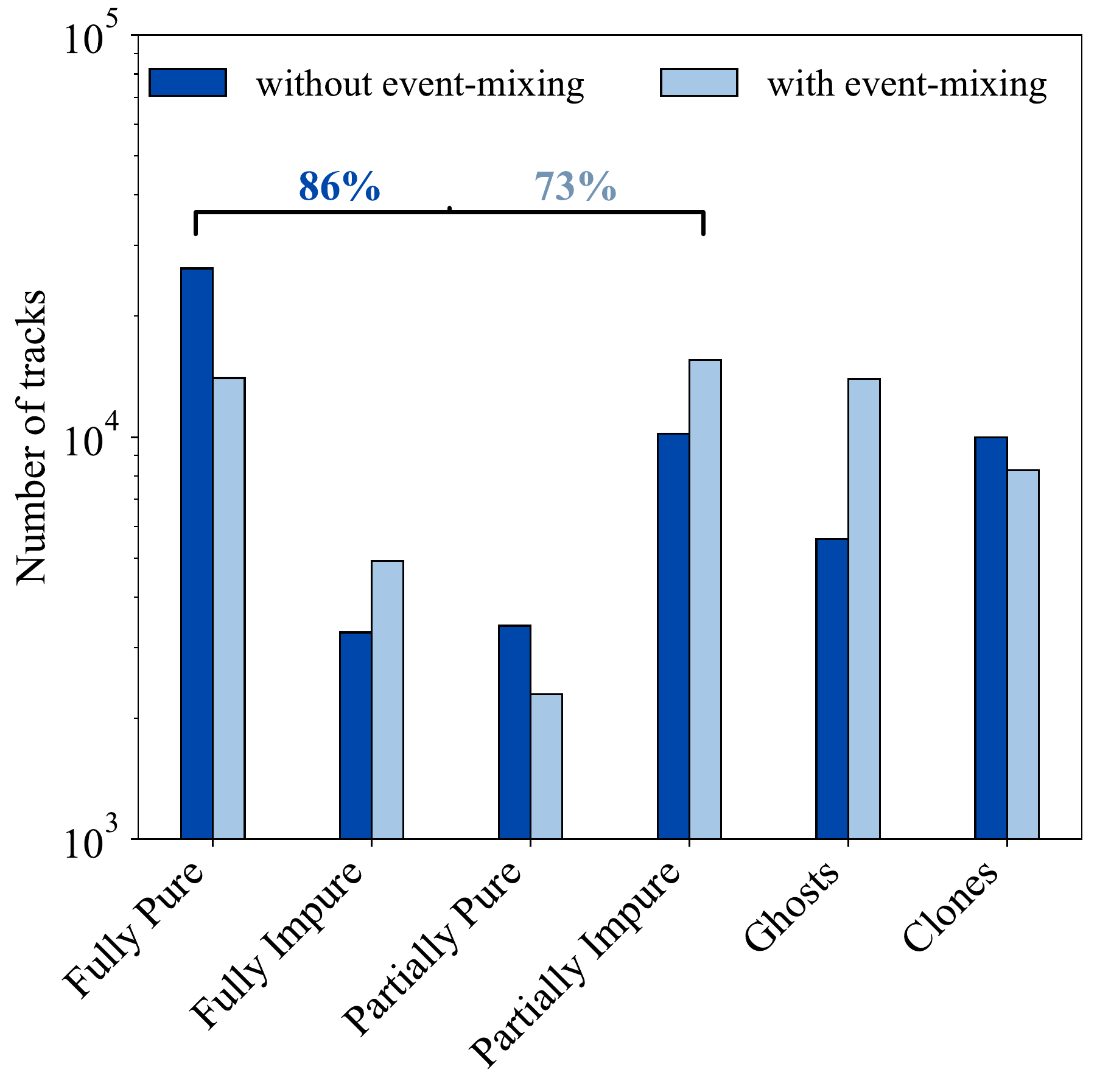}
    \caption{Histogram presenting the distribution of reconstructed tracks in each rank for the 50,124 simulated tracks in the 3~GeV/$c$ antiproton beam momentum simulation. The dark blue distribution shows the results for the 15,000 events processed one by one (no event-mixing),  and the light blue distribution shows the results for the setup where hits from 4 successive events are processed at once (mimicking the effect of event-mixing). The percentage shown represents the fraction of reconstructed tracks that are \textit{found} (\textit{i.e.}, having ranks between 1 and 4, see Section~\ref{sect:compari} for details).}
    \label{fig:qamix}
\end{figure}

\subsection{Towards high-luminosity experiments}
\label{sect:phaseII}

With the foreseen interaction rates and the relatively slow readout of the STT, one expects a large overlap in events for the \panda\ high-luminosity mode. In theory, given the maximum drift time of the tubes (250 ns) and the expected 20 MHz interaction rate (two successive events separated on average by 50 ns), hits belonging to 4 to 5 independent events might overlap and make the reconstruction process harder due to the larger number of overlapping tracks. In practice, taking into account that events will be processed in bursts of no less than 2,000 ns and the 400 ns beam gap, we expect that on average 32 events will mix at 20 MHz (3.2 at 2 MHz).

The burst building scheme of \panda\ implies that the data flow from the STT detector will be grouped in a time-based manner where track reconstruction algorithms not only have to find the tracks in real-time but also group these tracks into separate events. Therefore, any in-situ track reconstruction algorithms must account for the time-stamps of the hits as an additional dimension (see \cite{Papenbrock:2020znv} for a recent review on time-based reconstruction algorithms). Performing track-reconstruction in a time-based manner has a supplemental computational cost which can be, when all detectors' information occurring within 2,000 ns is processed, up to $\approx$ 1.5 times higher than the typical event-based time reconstruction \cite{Regina:2019dvn}.

In this Section, we explore the impact of event mixing on the \textsc{lotf} performance discussed in the previous section. To mimic these effects, we use the 15,000 events generated with an antiproton beam momentum of 3 GeV/$c$ and re-arrange them into 3,750 data sets of 4 events each. Note that this is an idealization assuming that the data collected is in an event-based manner, \textit{i.e.}, we do not consider the hits time-stamps as \textsc{lotf} is not yet able to deal with this 4$^{\rm th}$ data-dimension.  Yet, this setup enables a fair initial investigation of the ability of our algorithm to disentangle overlapping tracks in more complex data sets.

 Figure~\ref{fig:evtcomp} shows an example of track reconstruction based on one data set encompassing hits from 4 successive events. As can be seen in this figure, the number of overlapping tracks greatly increases in this setup, which significantly impacts the overall quality of the track reconstruction.

\begin{sloppypar}

To assess the performance of the \textsc{lotf} algorithm for these more complex cases, we use the metric described in Section~\ref{sect:compari}. Figure~\ref{fig:qamix} compares the distribution of track ranks for the set of reconstructed tracks based on the original simulation with a beam momentum of 3~GeV/$c$ with 15,000 events and based on the new setup composed of 3,750 data sets of 4 events each. The percentage of \textit{found} tracks decreases from 86\% to 73\% for the setup mimicking the impact of event-mixing. The number of \textit{Fully} and \textit{Partially Pure} tracks decreases while the number of \textit{Fully} and \textit{Partially Impure} tracks increase. This is expected since, as more and more simulated tracks overlap, the reconstructed tracks are more likely to include hits from several tracks. 

The number of \textit{Clones} remains fairly constant and we note a significant increase in the number of \textit{Ghost} tracks.  \textit{Ghost} tracks are the consequence of the algorithm combining too many hits belonging to different simulated tracks. These tracks have not been matched to any simulated tracks, and emphasize the larger complexity of accurately reconstructing individual tracks when the likeliness that tracks overlap is higher. We note that accounting for the hits time-stamps can greatly improve the efficiency of the track reconstruction and significantly reduce the number of \textit{Ghost} tracks because it adds an additional constraint when selecting a group of hits that might belong to the same track (see \textit{e.g.}, \cite{Regina:2020ctd}). Extending \textsc{lotf} implementation to handle the time dimension is one of our priorities.
\end{sloppypar}

Overall, a loss of performance is expected given the more complex nature of the event-mixing case, yet the algorithm still performs well in identifying a large fraction of the tracks with decent accuracy.  These results are promising for future \panda\ experiments operating at interaction rates of up to 20 MHz.

\subsection{Speed of the algorithm}
\label{sect:time}

\begin{figure}[!htbp]
    \centering
    \includegraphics[width = 0.45\textwidth]{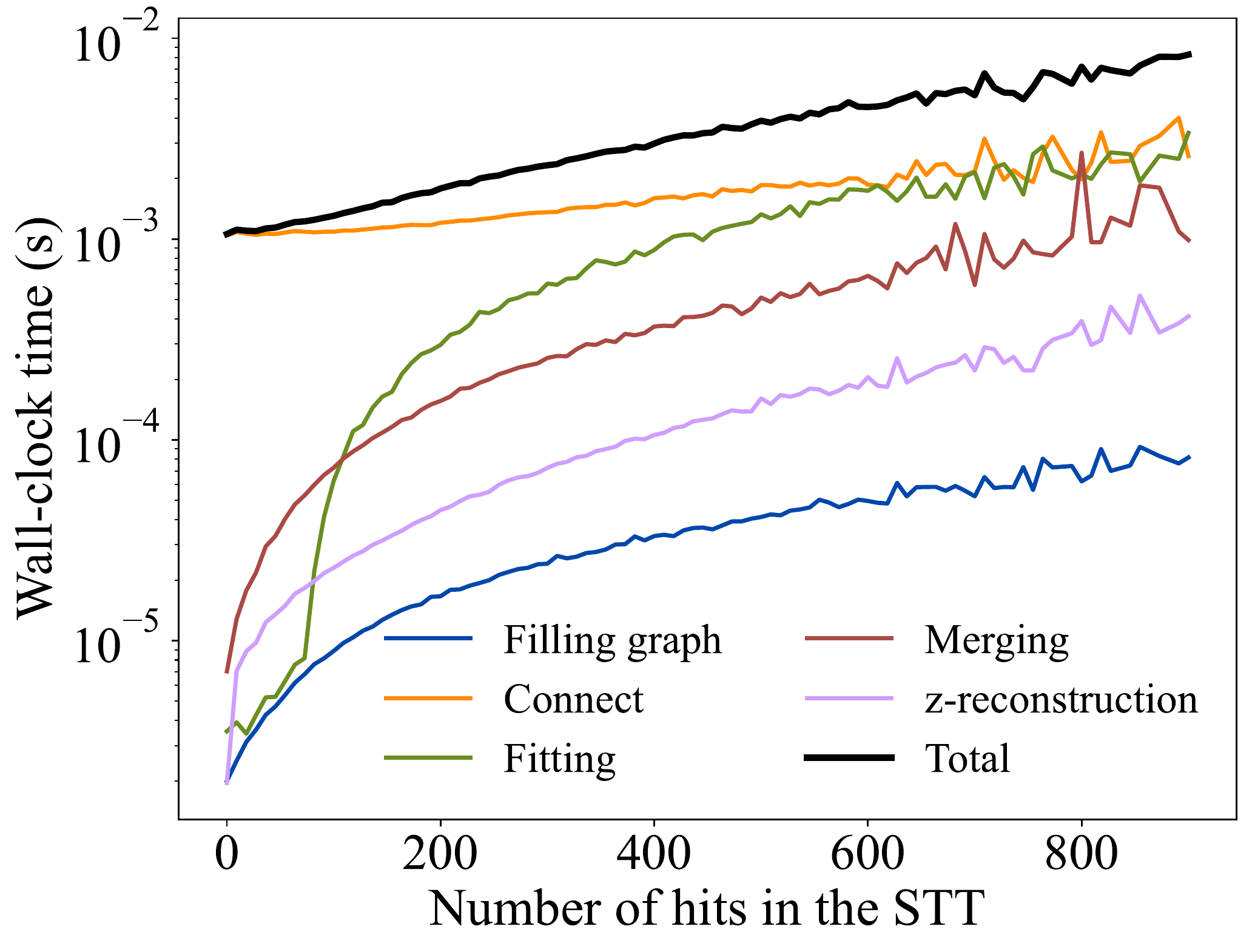}
        \includegraphics[width = 0.45\textwidth]{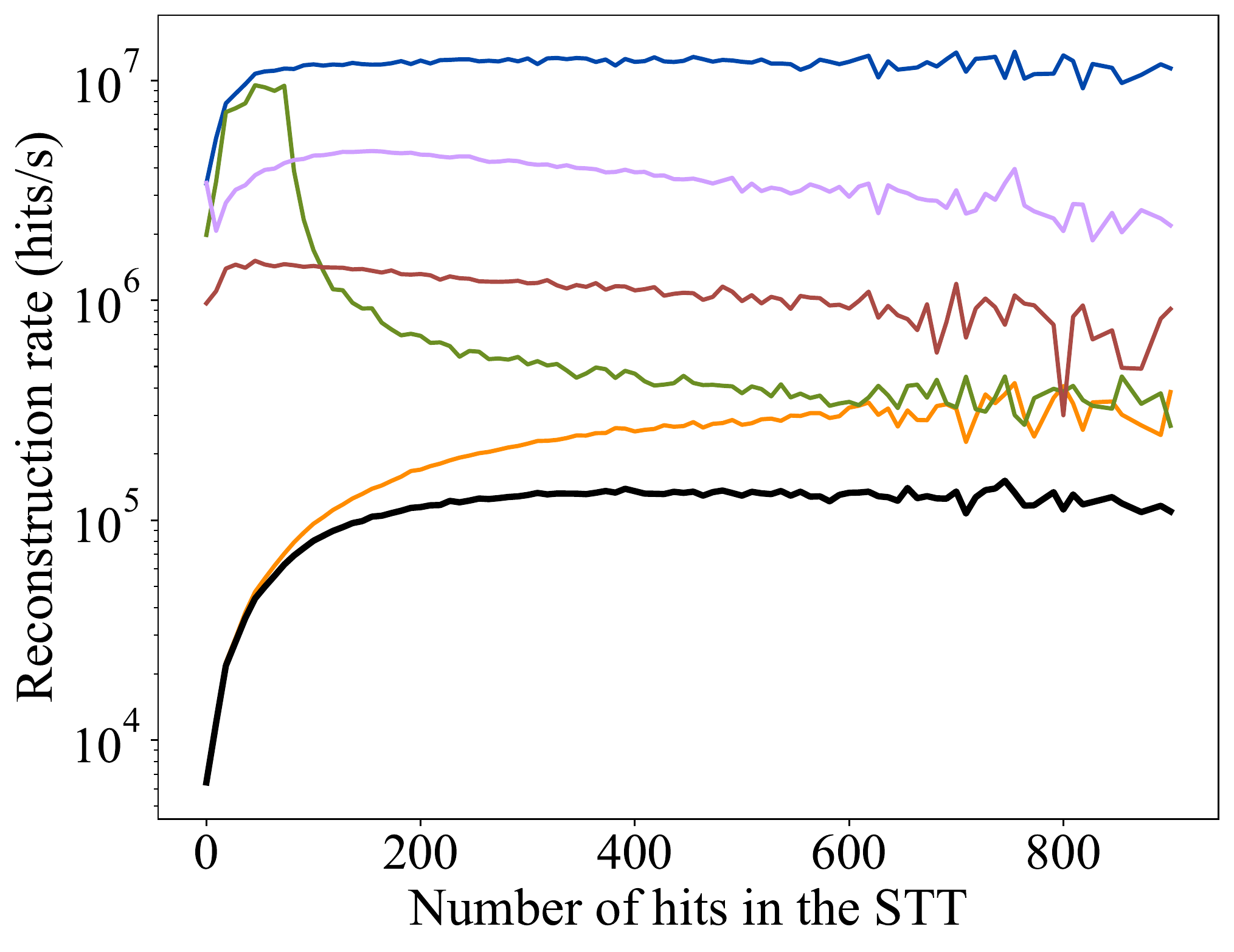}
    \caption{The evolution of the wall-clock time (top panel) and reconstruction rate (in hits/s, bottom panel) as a function of the number of STT hits in the 33,750 events processed in this work. We decomposed the total time based on the five major processing steps performed by \textsc{lotf} (see details in Section~\ref{sect:time}). The connect phase takes more than 50\% of the total reconstruction time for data sets with less than 300 hits. The fitting phase contribution is negligible for data sets with less than 100 STT hits, but becomes more and more significant as the number of hits increases, reaching 30\% of the total time for data sets with more than 300 hits. Overall, the average reconstruction rate is 67,849 hits/s ($\sim$107 tracks/s assuming 21 hits per track) and remains constant as the number of hits in the STT increases. } 
    \label{fig:timing}
\end{figure}
\renewcommand{\arraystretch}{2}

% \begin{table}[!htbp]
%     \centering
%     \begin{tabular}{c|c}
%        Processing step &  Time as function of $n_{\rm hits}$ (s) \\
%        \hline \hline
%         Filling graph &  $4.16\times10^{-8}\times (n_{\rm hits})^{1.10} + 2.20\times10^{-6} $\\
%         Connect &  $1.07\times10^{-7}\times (n_{\rm hits})^{1.42} + 1.02\times10^{-3}$\\
%         Fitting &  $2.26\times10^{-7}\times (n_{\rm hits})^{1.39} + 1.26\times10^{-13} $\\
%         Merging &  $1.01\times10^{-7}\times (n_{\rm hits})^{1.36} + 1.80\times10^{-5} $\\
%         z-reconstruction &  $1.81\times10^{-8}\times (n_{\rm hits})^{1.43} + 8.84\times10^{-6} $\\
%         \hline
%         Total &  $3.75\times10^{-7}\times (n_{\rm hits})^{1.43} + 1.04\times10^{-3} $\\

%     \end{tabular}
%     \caption{\textbf{The expected time dependence on the number of hits for each processing step in the \textsc{lotf} algorithm. Each equation has been obtained by fitting an equation of the form $ t = A \times (n_{\rm hits})^B + C$ to each curve shown in the top panel of Figure~\ref{fig:timing}}.} 
%     \label{tab:timedep}
% \end{table}

Here, we assess the speed performance of the \textsc{lotf} algorithm. We test our method using a laptop with a single core of an Intel i9-11900H CPU at 2.50~GHz. We measure the time and reconstruction rate (hits/s) for all events and explore its evolution as a function of the number of hits per data set processed. We consider the timings of all the 33,750 events (30,000 single events plus 3,750 data sets with 4 events concatenated) processed in this work. The total number of tracks considered is 179,523, and each track holds on average 21 hits. Over the 33,750 data sets considered, only 424 ($\approx$1\%) have more than 500 STT hits. These cases typically correspond to the data sets with event-mixing used in Section~\ref{sect:phaseII}. To have a complete assessment of the computational footprint of our algorithm, we decompose the total time of the track reconstruction in several processing steps:

\begin{itemize}
    \item ``filling graph": the time needed to fill the STT graph with the collected hits;
    \item ``connect": the time needed to complete the \textit{connect} phase; 
    \item ``fitting": the time needed to complete the \textit{fitting} phase;
    \item ``merging": the time needed to complete the \textit{merging} phase;
    \item ``z-reconstruction": the time needed to perform the z-reconstruction.
\end{itemize}

\noindent In all time measurements presented hereafter, we do not consider input/output overheads. The top panel of Figure~\ref{fig:timing} shows the evolution of the average wall-clock time for each processing step as a function of the number of hits in the STT per data set. 

%\textbf{Additionally, we provide in Table~\ref{tab:timedep} the expected time dependence on the number of STT hits ($n_{\rm hits}$) for each processing step. We obtained each equation by fitting a power law of the form $t = A \times (n_{\rm hits})^B + C$, where A, B, and C are the fitted parameters. We chose a power law because we found that it provides the best representation for all the curves observed in the top panel of Figure~\ref{fig:timing}.} 

The most time-consuming phase is the connect phase where hits are connected one by one as a function of their spatial and neighborhood relations. In particular, this phase takes more than 50\% of the total reconstruction time for data sets with less than 300 hits. The contribution of the fitting phase is not significant for events with less than 100 hits, but rapidly increases as the number of hits per data becomes larger ($\sim$30\% of the total time for $n_{\rm hits} > 300$). This is somewhat expected as data sets with a larger number of STT hits likely have a larger number of overlapping tracks as well, such that an increasing amount of time must be spent during this phase to reconstruct these complex tracks. Finally, we note that the merging, filling graph, and z-reconstruction time are the least time-consuming processing steps, mostly because they involve very simple tasks. 

In \cite{babai20}, the authors estimate the complexity of their method to be $O(nlog(n))$. We find here that the total reconstruction time is fairly consistent with this function. Additionally, we tried to empirically estimate the time dependence of each processing step on the number of STT hits by fitting a power law to the different curves observed in Figure~\ref{fig:timing}. We find all processing steps have a computational complexity close to $O(n^{1.5})$, with the exception of the filling graph whose complexity is $O(n)$. We must note that these values should only be considered as rough estimates. Indeed, empirically estimating the asymptotic behavior of our method for large $n$ is complex because we are dealing with modest values of $n$ ($< 1000$).

The lower panel of Figure~\ref{fig:timing} which presents the evolution of the reconstruction rate as a function of the number of STT hits shows similar trends. Interestingly, the reconstruction rate of all processing steps remains fairly constant when the number of hits considered is larger than 200. This scaling behavior is promising for future developments that will aim to limit our method's computational footprint using an appropriate parallelization scheme (see discussion hereafter). 

\begin{sloppypar}

The average processing time per data set is about 1.5~ms for the case where events are processed one by one, and 2.3~ms for the 3,750 data sets where 4 events are processed simultaneously. The average reconstruction rate is  67,849~hits/s or $\sim$107~tracks/s assuming an average of 21 hits per track based on our experiment. This rate is lower for data sets with a small number of hits (between 10,000 and 50,000~hits/s for $10 < n_{\rm hits}< 100$), but reaches a plateau at 124,345~hits/s for data sets with a number of STT hits larger than 200. As a comparison, running with the lowest and highest luminosities, \panda\ is expected to generate around 160~Mhits/s and 1,600~Mhits/s in the STT for the high-resolution and high-luminosity modes, respectively. Hence, an improvement of more than a factor of 3 and 4 orders of magnitude is required to perform efficient online track reconstructions for both operation modes.

We are confident that future developments will enable us to get closer to the requirements set by the expected interaction rates of \panda. Currently, \textsc{lotf} does not include any parallelization scheme. The most time-consuming phase, the connect phase, involves simple and separable series of tasks that can be easily distributed over several processes. Parallelizing the fitting phase would be slightly more complex as all operations are not entirely separable, yet, decent gains should still be reachable by distributing smaller operations (\textit{e.g.}, the selection of the best track continuation based on the fit prediction). Additionally, a GPU-based (\textit{e.g.} as in \cite{Schumann:139096, herten2015gpu}) or FPGA-based implementation \cite{Liang2017} could help to further increase the average reconstruction rate to meet the requirements for the PANDA in-situ track reconstruction. The latter approach is particularly promising as the authors report an improvement of up to 3 orders of magnitude compared to classical CPU-based approaches.  Alternatively, since each event can be reconstructed independently, one could employ this method on a machine with 100 (1,000) cores and meet the requirements set by the foreseen interaction rates of the high-resolution (high-luminosity) operational mode.

\end{sloppypar}

% Our method is significantly faster, with an average of 0.037 seconds per event and a speed-up factor of 145 compared to the algorithm of \bab. These results are promising to enable fast-decision making in online event selection, yet further improvements are required to match the 20 MHz data acquisition rate of \panda. We note that several parts of our algorithm can be distributed over several processes to speed up this reconstruction. In particular, the operations performed during the \textit{connect} phase, which looks for track extremities, and the \textit{merging} phase are parallelizable series of tasks. The \textit{fitting} phase would be slightly more complex as all operations are not entirely separable, yet, a decent gain should still be reachable by distributing smaller operations (\textit{e.g.}, the selection of the best track continuation based on the fit prediction).  Using a GPU parallelization scheme for such operations is particularly attractive to reach an even lower computational time per event because GPUs are suited to complete redundant tasks efficiently. We plan to explore this in future work. \\
%(0.02471842775093333, 0.02688037497213333, 0.08290973663454546)

\section{Summary}
\label{sect:conc}

\begin{sloppypar}
Designing fast and efficient track reconstruction algorithms is crucial to meet the requirements of modern particle detectors operating at very high interaction rates. In this work, we presented the LOcal Track Finder (\textsc{lotf}) algorithm that performs fast track reconstructions using the data collected by the Straw Tube Tracker embedded in the upcoming \panda\ experiment. The functioning of our algorithm uses a local approach that connects single isolated hits to form tracks. Further, it builds upon a parametric linear fitting method to refine the tracks in regions where several particle trajectories overlap. Additionally, we utilize the virtual nodes system introduced in  \citet{babai16, babai20} to perform the z-reconstruction.  Our approach does not depend on the drift time of the tubes and only requires the STT geometry, including the neighborhood relation of all tubes (similar to \cite{ADINETZ2014, Regina:2019dvn, babai20}).
\end{sloppypar}

\begin{sloppypar}

In Section~\ref{sect:phaseI}, we generated 30,000 events (107,605 particle trajectories with at least 6 STT hits) using an antiproton beam momentum of 3 and 15~GeV/$c$ to assess the performance of our algorithm. We compare our method to the method of \citet{babai20} and to the method implemented in the \textsc{PandaRoot} software (\textsc{BarrelTrackFinder}) using the standard \panda\ track quality assessment metric.  We showed that, in the low and high beam momentum simulations, \textsc{lotf} has (1) respectively 86\% and 84\% of \textit{found} tracks (\textit{i.e.}, tracks having a rank between 1 to 4), (2) the largest number of \textit{Fully Pure} tracks (rank 1, perfect reconstructions), (3) the lowest amount of \textit{Ghosts} and \textit{Clones} (ranks 5 and 6, incorrect reconstructions), and (4), is faster by a factor of $\sim$13 and 940 compared to \textsc{BarrelTrackFinder} and \bab, respectively. Additionally, we emphasized that our z-reconstruction approach leads to an average z-error of $1.80^{+9.79}_{-10.38}$ cm for all tracks which is promising for extracting the particles' longitudinal momentum component.
\end{sloppypar}

% We tested the ability of our algorithm to extract a first estimate of the particle's transverse momentum component on-the-fly for the $\sim$48,000 reconstructed tracks that have a rank of 1 or 2, and transited through a sufficient portion of the STT volume ($>$ 16 STT hits). %Our approach enables a relatively precise (relative error less than 10\%) reconstruction of $p_T$ for particles with a transverse momentum lower than 0.3 GeV/$c$ only. 

In Section~\ref{sect:phaseII}, we mimicked the effects of event-mixing to further explore the foreseen performance of our algorithm for experiments using the \panda\ high-luminosity mode (20 MHz interaction rate, involving a significant overlap between events). Using the simulation with 15,000 events simulation generated with an antiproton beam momentum of 3~GeV/$c$, we created 3,750 data sets combining each of the STT hits from 4 successive events. We showed that the percentage of \textit{found} tracks decreases (86\% to 73\%) in the setup mimicking event-mixing. Yet, \textsc{lotf} still performs well in identifying a large fraction of the tracks with a decent accuracy according to the quality assessment metric used in this work,  which is promising for dealing with such complex event-mixing data sets.

In Section~\ref{sect:time}, we scrutinized the time and processing rate of our method. We obtained an average of 1.5~ms per data set for data sets composed of a single event and 2.3~ms per data set for data sets composed of 4 successive events. The average processing rate is about 68,000 hits/s for both cases. The current efficiency and raw computational speed (no parallelization included) are promising in enabling a fast in-situ event selection with \panda. To make this method a robust state-of-the-art algorithm for \panda, we are planning several improvements. We aim at decreasing the current processing time per event by 2 orders of magnitudes by optimizing and parallelizing the phases having a large computational footprint (\textit{e.g.}, the \textit{connect} and \textit{fitting} phases) using GPUs. Then, in practice, since each event can be reconstructed independently, another improvement of two and three orders of magnitude  should be achievable by employing this method on a machine with 100 (1,000) cores, and therefore, enable an efficient in-situ event selection for experiments working with the \panda\ high-resolution (high-luminosity) mode. 

%Additionally, using a smaller subset of the STT nodes to perform the track reconstruction (\textit{e.g.}, similar to \cite{ADINETZ2014}),  might help to further speed up our method while keeping a similar reconstruction efficiency. \\

Finally, in the future, we will aim to include the data collected by the MVD \citep{pandamvd} and EMC \citep{pandaemc} detectors in the reconstruction to provide further constraints on the particle trajectories inside and outside of the STT volume. This effort will enable us to improve the efficiency of our algorithm in identifying and reconstructing overlapping particle trajectories. However, we must ensure that these improvements do not significantly impact the current speed of the reconstruction process.

\section*{Acknowledgements}
\begin{sloppypar}

This paper is based on research developed in the DSSC Doctoral
Training Programme. The authors would like to thank the referees whose comments greatly helped improve the quality of this manuscript. SG thanks Mohammad Babai for sharing his code, and for the very useful discussions throughout this work. SG also thanks the \panda\ collaboration for the development of the \textsc{PandaRoot} software. SG acknowledges the use of the python packages  \textsc{numpy} \cite{numpy}, \textsc{matplotlib} \cite{matplotlib}, and \textsc{panda} \cite{panda}.
\end{sloppypar}
\section*{Declaration}
\subsection*{Funding}
This work was funded by the Centre for Data Science and Systems' Complexity, University of Groningen.

\subsection*{Conflicts of interest/Competing interests}
Not applicable
\subsection*{Availability of data and material}
The simulated data sets underlying this paper are available upon request to the corresponding author. 
\subsection*{Code availability}
The code of the \textsc{lotf} algorithm is available at \url{https://github.com/sgazagnes/lotf}.

% \begin{figure}
% % Use the relevant command for your figure-insertion program
% % to insert the figure file.
% % For example, with the option graphics use
% \resizebox{0.75\textwidth}{!}{%
%   \includegraphics{leer.eps}
% }
% % If not, use
% %\vspace{5cm}       % Give the correct figure height in cm
% \caption{Please write your figure caption here}
% \label{fig:1}       % Give a unique label
% \end{figure}
% \begin{figure*}
% % Use the relevant command for your figure-insertion program
% % to insert the figure file. See example above.
% % If not, use
% \vspace*{5cm}       % Give the correct figure height in cm
% \caption{Please write your figure caption here}
% \label{fig:2}       % Give a unique label
% \end{figure*}
% \begin{table}
% \caption{Please write your table caption here}
% \label{tab:1}       % Give a unique label
% % For LaTeX tables use
% \begin{tabular}{lll}
% \hline\noalign{\smallskip}
% first & second & third  \\
% \noalign{\smallskip}\hline\noalign{\smallskip}
% number & number & number \\
% number & number & number \\
% \noalign{\smallskip}\hline
% \end{tabular}
% % Or use
% \vspace*{5cm}  % with the correct table height
% \end{table}
%
% BibTeX users please use
% \bibliographystyle{unsrt}
% \bibliography{Bibli.bib}
%

\printbibliography
\appendix
\label{app:app}
\section{Algorithms}
In this section, we present the code of the algorithms used in the present work. 
\subsection{The \textit{connect} phase}
\label{app:connect}

\begin{algorithm}
\caption{Pseudo-code of the \textit{connect} phase}
\label{alg:connect}
\begin{algorithmic}[1]
    \STATE \textbf{procedure }\textsc{FindTracklets}(${\rm STT_{\rm graph}}$)
\begin{ALC@g}
{\small
\STATE $hit\_queue  \gets \textsc{FindStartingHits}({\rm STT_{\rm graph}})$
\FORALL {hit in $hit\_queue$}
\STATE $CurHit \gets \text{hit}$
\STATE PathCandidate $Cand \gets \textsc{init}(CurHit)$
%\STATE $CurrentHit \gets \text{mark as visited}$
\STATE $CurDir \gets 0$
\STATE $Cond \gets true$
\STATE $neighbors \gets \textsc{FindNeighbors}(CurHit)$
\WHILE{$Cond$}
\IF{$neighbors.\textsc{size()} = 1$}
\STATE $Cand.\textsc{insert}(neighbors[0])$
%\STATE $neighbors[0] \gets \text{mark as visited}$
\STATE $CurDir \gets \textsc{FindDir}(CurHit, neighbors[0])$
\STATE $CurHit \gets neighbors[0]$
\STATE $neighbors \gets \textsc{FindNeighbors}(CurHit)$
\ELSIF{$size(neighbors) > 1$}
\STATE $\textsc{SortPerLayer}(neighbors, upL, downL, sameL)$
\IF{$(upL.\textsc{size()} > 0$ \textbf{and} $downL.\textsc{size()} > 0$ \\ \textbf{and} ($sameL.\textsc{size()} > 0)$ \textbf{or} $CurDir = SAME$)}
\STATE $Cond \gets false$
\ELSIF{$(size(upL) > 0$ \textbf{and} $downL.\textsc{size()} > 0$}
\IF{$CurDir = UP$}
\STATE $neighbors.\textsc{remove}(downL)$
\ELSIF{$CurDir = DOWN$}
\STATE $neighbors.\textsc{remove}(upL)$
\ENDIF
\IF{$Cond = true$ \textbf{and} \textsc{AreConnected}($neighbors$)}
\STATE $Cand.\textsc{insert}(neighbors)$
\STATE $CurDir \gets \textsc{FindDir}(CurHit, neighbors)$
\STATE $neighbors \gets \textsc{FindNeighbors}(neighbors)$
\STATE $CurHit \gets neighbors[-1]$
\ENDIF
\ENDIF
\ELSE
\STATE $Cond \gets false$
\ENDIF
\ENDWHILE
\STATE $Cand.neighbors.\textsc{insert}(neighbors)$
\IF{\textsc{OnLayerLimit}($Cand.headNode$) \textbf{and} \textsc{OnLayerLimit}($Cand.tailNode$)  \textbf{and} $neighbors.\textsc{size()}$ = 0}
\STATE $Cand.status \gets finished$
\ELSE
\STATE $Cand.status \gets ongoing$
\ENDIF
\ENDFOR

}
\end{ALC@g}
\STATE \textbf{end procedure}
\end{algorithmic}
\end{algorithm}

In Section~\ref{sect:connect}, we briefly presented the \textit{connect phase}. In this appendix, we present and detail the corresponding pseudocode in Algorithm~\ref{alg:connect}. The function \textsc{FindStartingHits} looks for the edges of the track by extracting all nodes that belong to the inner- or outermost layer of the STT (the ``limit layers"). Every time we find such a node,  we define a new track candidate using the structure \textit{PathCandidate} and set the current node as the tail of the track. Neighbors of this node are extracted and stored in the \textit{neighbors} vector. We use a variable \textit{CurDir} to determine the current orientation of the track, based on the layer of the nodes that have been added so far (layer 0 being the closest to the beam-target interaction point). The \textit{CurDir} variable is either \textit{UP}, \textit{DOWN}, or \textit{SAME} depending on whether the layer index difference of the two most recently added nodes is $1$, $-1$, or $0$, respectively.    

We connect neighboring activated tubes iteratively to a track using a loop that successively looks for the next available neighbors around the most recently added node. If we have a single activated tube, it is automatically connected to the track. Otherwise, we use the procedure \textsc{SortPerLayer} to sort all the available neighboring nodes based on their layer index. We then add only neighbors that are consistent with the track direction and are adjacent to each other  (function \textsc{AreConnected}). If the layer configuration of all neighbors is too complex (\textit{i.e.} we have neighbors on all different layers and the track direction is unclear), we pause the track reconstruction and store the current list of neighbors.  Then, the track reconstruction is resumed during the \textit{fitting} phase to resolve these complex cases.  

The track reconstruction also stops when we have no more neighboring nodes in sight. For the latter case, we test whether the current track looks complete by testing if its tail node (the first node added) and head node (the last node added) are on layer limits. This criterion assumes that a track transiting through the entire STT volume or circling in it is finished. If a track has no more neighboring nodes but one of its extremities ends in the middle of the STT, we flag it as \textit{ongoing} for later investigation. Indeed, there might exist another track candidate, not directly neighboring, that it can be connected to.

\subsection{The \textit{fitting} phase}
\label{app:fit}

\begin{algorithm}
\caption{Pseudo-code of the \textit{fitting} phase}
\label{alg:fitting}
\begin{algorithmic}[1]
    \STATE \textbf{procedure }\textsc{FitNextHit}(${\rm STT_{\rm graph}}$, PathCandidate $tracklets$)
\begin{ALC@g}
{\small
\FORALL {$track$ in $tracklets$ with $track.status = ongoing$}
\STATE $neighbors \gets track.neighbors$
\STATE  $DistDir \gets $ New 2-D tupple array
\STATE $Cond \gets true$
\WHILE{$Cond$}
\FORALL {hit in $neighbors$}
\STATE $Anchors \gets CurTrack.\textsc{ExtractLast3Anc}()$
\STATE $xInterp, yInterp \gets \textsc{FitXYLine}(Anchors)$
\STATE $CurDir \gets \textsc{FindDir}(Anchors)$
\STATE $xPred, yPred \gets \textsc{HitPos}(xInterp, yInterp)$
\STATE $hitDist \gets \textsc{ComputeDist}(hit, xPred, yPred)$
\STATE $hitDir \gets \textsc{CompareDir}(CurDir, hit)$
\STATE $DistDir.\textsc{insert}(hitDist, hitDir)$
\ENDFOR
\STATE $bestHit \gets \textsc{FindBestHit}(DistDir)$
\IF{$bestHit = -1$}
\STATE $Cond \gets false$
\ELSE
\STATE $hitTrack \gets \textsc{FindTrackOf}(bestHit)$
\IF{$hitTrack \neq -1$ \textbf{and} $\textsc{CheckMerging}(track, hitTrack) = 1$}
\STATE \textsc{AddTrackForMerging}($track$, $hitTrack$)
\STATE $Cond \gets False$
\ELSE
\STATE $track.\textsc{insert}(bestHit)$
\STATE $neighbors \gets \textsc{FindNextNeighbors}(bestHit)$
\ENDIF
\ENDIF
\IF{$neighbors.\textsc{size()}$ = 0}
\STATE $Cond \gets false$
\ENDIF
\ENDWHILE
\IF{\textsc{LayerLimit}($track.headNode$) \textbf{and} \textsc{LayerLimit}($track.tailNode$)}
\STATE $track.status \gets finished$
\ENDIF
\ENDFOR
}
\end{ALC@g}
\STATE \textbf{end procedure}
\end{algorithmic}
\end{algorithm}

Algorithm~\ref{alg:fitting} details the steps described in Section~\ref{sect:fitting}. For all the tracks flagged as \textit{ongoing}, we use a local fitting approach based on a system of \textit{anchor} nodes to look for the next best node to add to the track. The function \textsc{ExtractLast3Anc} recovers the last three anchors in the track that are used to fit the parametric equation system for the x and y coordinates (function \textsc{FitXYLine}). The variables \textit{xPred} and \textit{yPred} are the predicted coordinates of the next nodes according to the parametric equations. For all the neighboring nodes, we compute the distance (labeled \textit{hitDist}) between the node position and the predicted coordinates of the next hit using the function \textsc{ComputeDist}. Additionally, we use the anchors to determine the track direction (\textit{CurDir}) which is derived based on the layers the anchors belong to (similar to the method used during the \textit{connect} phase). Once we tested all the hits, we look for the best possible fit in the list by taking the node with the minimal distance while having a consistent direction with respect to the \textit{CurDir} variable. This is done using the function \textsc{FindBestHit}. As detailed in Section~\ref{sect:fitting}, we set a criterion on the maximum distance acceptable such that the function returns $-1$ if no neighboring nodes are sufficiently close to the predicted node position.

When a good node is found, if it already belongs to another track, we check whether these tracks should be merged using the function \textsc{CheckMerging} which computes the intersection angle between two tracks. If the \textsc{CheckMerging} test is successful (see Section~\ref{sect:fitting}), the fitting phase for this particular track is stopped, and both tracks are flagged as \textit{ToMerge} such that no more hits are added.

On the other hand, if \textsc{CheckMerging}  returns false, or if the best node found does not belong to any track,  we connect this node to the current track. We repeat the steps above updating the list of neighbors to include the neighbors of the most recently added tube. The algorithm continues until the current track is flagged for merging, or until the list of next neighbors is empty.

\subsection{z-reconstruction}
\label{app:zrec}

In this section, we detail the pseudo-code for the procedures used for the z-reconstruction in Algorithm~\ref{alg:zrec}. The \textsc{CorrectSkewedXY} procedure is the function used to estimate the exact hit position along the tube (see Figure~\ref{fig:zskew}). The procedure \textsc{InterpolateZCoord} is the function used at the end of the track reconstruction to fit a parametric line in the z-direction and re-estimate consistently the z coordinates of all the nodes in the track. The functioning of the two procedures is detailed in Section~\ref{sect:zrec}. 

\begin{algorithm}
\caption{Pseudo-code of the z-reconstruction step}
\label{alg:zrec}
\begin{algorithmic}[1]
\STATE \textbf{procedure }\textsc{CorrectSkewedXY}(${\rm STT_{\rm graph}}$, PathCandidate $track$, GridNode $hitToInsert$)
\begin{ALC@g}
{\small
\IF{$hitToInsert$ is a virtual node}
\STATE GridNode $lastVirt \gets \textsc{ExtractVirt}(track)$
\STATE $xDir \gets hitToInsert.x - lastVirt.x$
\STATE $yDir \gets hitToInsert.y - lastVirt.y$
\FORALL {$anchors$ in $track$ added after $lastVirt$} 
\STATE $anchor.x, anchor.y \gets~\textsc{Intersect}(xDir, yDir)$
\STATE $anchor.z \gets \textsc{EstimateZCoord}(anchor)$
\ENDFOR
\ENDIF
}
\end{ALC@g}
\STATE \textbf{end procedure}
\vspace{1cm}
\STATE \textbf{procedure }\textsc{InterpolateZCoord}(PathCandidate $track$)
\begin{ALC@g}
{\small
\STATE $zArray \gets \textsc{ExtractAllAnchorsWithZCoord}(track)$
\IF{$track.Dir = InnerToOuter$}
\STATE $zArray.insertAtFirstPos(0)$
\ELSIF{$track.Dir = OuterToInner$}
\STATE $zArray.insertAtLastPos(0)$
\ENDIF
\STATE $zInterp \gets \textsc{FitZLine}(zArray)$
\FORALL {nodes in $track$} 
\STATE $nodes.z \gets \textsc{ComputeZFromInterp}(zInterp)$
\ENDFOR
}
\end{ALC@g}
\STATE \textbf{end procedure}
\end{algorithmic}
\end{algorithm}

\end{document}